\documentclass[aoas,preprint]{imsart}

\usepackage{amsthm,amsmath}
\usepackage[authoryear]{natbib}
\RequirePackage[colorlinks,citecolor=blue,urlcolor=blue]{hyperref}

\usepackage{graphicx,psfrag,epsf}
\usepackage{enumerate}
\usepackage{url}
\usepackage{color}
\usepackage[T1]{fontenc}
\usepackage{subcaption}
\usepackage{amssymb}
\usepackage{mathtools}
\usepackage{booktabs}
\usepackage{color}
\usepackage{multicol}
\usepackage{floatrow}
\usepackage{todonotes}
\usepackage{placeins}

\presetkeys{todonotes}{color=blue!30,size=\small}{}

\newfloatcommand{capbtabbox}{table}[][\FBwidth]

\graphicspath{{figures/}}
\DeclareGraphicsExtensions{.pdf,.jpeg,.jpg,.png}

\usepackage{pgfplots}
\pgfplotsset{compat=1.13}
\usepgfplotslibrary{colorbrewer}
\pgfplotsset{cycle list/Set1-4}

\arxiv{1803.00113}

\startlocaldefs

\endlocaldefs

\makeatletter

\g@addto@macro{\UrlBreaks}{\UrlOrds}

\begin{document}

\begin{frontmatter}

\title{Approximate Inference for Constructing Astronomical Catalogs from Images}
\runtitle{Astronomical Cataloging}

\author{\fnms{Jeffrey} \snm{Regier}\corref{}\texorpdfstring{$^1$}{1}\ead[label=e1]{jregier@eecs.berkeley.edu}},
\address{
Jeffrey Regier\\
Department of Electrical Engineering\\and Computer Sciences\\
University of California, Berkeley\\
465 Soda Hall\\
Berkeley, CA 94720\\
USA\\
\printead{e1}}
\author{\fnms{Andrew C.} \snm{Miller}\texorpdfstring{$^2$}{2}\ead[label=e2]{am5171@columbia.edu}},
\address{
Andrew C. Miller\\
Data Science Institute \\
Columbia University \\
3227 Broadway \\
New York, NY 10027 \\
USA\\
\printead{e2}}
\author{\fnms{David} \snm{Schlegel}\texorpdfstring{$^3$}{3}\ead[label=e3]{djschlegel@lbl.gov}},
\address{
David Schlegel\\
Lawrence Berkeley National Laboratory\\
1 Cyclotron Road\\
Berkeley, CA 94720\\
USA\\
\printead{e3}}
\author{\fnms{Ryan P.} \snm{Adams}\texorpdfstring{$^{4}$}{4}\ead[label=e4]{rpa@princeton.edu}},
\address{
Ryan P. Adams\\
Princeton University\\
Department of Computer Science\\
35 Olden Street\\
Princeton, NJ 08540\\
USA\\
\printead{e4}}
\author{\fnms{Jon D.} \snm{McAuliffe}\texorpdfstring{$^{1,5}$}{1,5}\ead[label=e5]{jon@stat.berkeley.edu}},
\address{
Jon D. McAuliffe\\
Department of Statistics\\
University of California, Berkeley\\
367 Evans Hall\\
Berkeley, CA 94720\\
USA\\
\printead{e5}}
\and
\author{\snm{Prabhat}\texorpdfstring{$^3$}{3}\ead[label=e6]{prabhat@lbl.gov}}
\address{
Prabhat\\
Lawrence Berkeley National Laboratory\\
1 Cyclotron Road\\
Berkeley, CA 94720\\
USA\\
\printead{e6}}
\\

\affiliation{%
University of California -- Berkeley\texorpdfstring{$^1$}{1},
Columbia University\texorpdfstring{$^2$}{2},\\
Lawrence Berkeley National Laboratory\texorpdfstring{$^3$}{3},\\
Princeton University\texorpdfstring{$^4$}{4}, and
The Voleon Group\texorpdfstring{$^5$}{5}.
}

\runauthor{J. Regier et al.}

\begin{abstract}
We present a new, fully generative model for constructing astronomical catalogs from optical telescope image sets. Each pixel intensity is treated as a random variable with parameters that depend on the latent properties of stars and galaxies. These latent properties are themselves modeled as random. We compare two procedures for posterior inference. One procedure is based on Markov chain Monte Carlo (MCMC) while the other is based on variational inference (VI). The MCMC procedure excels at quantifying uncertainty, while the VI procedure is 1000 times faster. 
On a supercomputer, the VI procedure efficiently uses 665,000 CPU cores to construct an astronomical catalog from 50 terabytes of images in 14.6 minutes, demonstrating the scaling characteristics necessary to construct catalogs for upcoming astronomical surveys.
\end{abstract}

\begin{keyword}[class=MSC]
\kwd[Primary ]{62P35}
\kwd[; secondary ]{85A35}
\end{keyword}

\begin{keyword}
\kwd{astronomy}
\kwd{graphical model}
\kwd{MCMC}
\kwd{variational inference}
\kwd{high performance computing}
\end{keyword}

\end{frontmatter}

\section{Introduction}
\label{sec:intro}
Astronomical surveys are the primary source of information about the universe beyond our solar system. They are essential for addressing key open questions in astronomy and cosmology about topics such as the life cycles of stars and galaxies, the nature of dark energy, and the origin and evolution of the universe.

\begin{figure}[hb]
\centering
\includegraphics[width=2in]{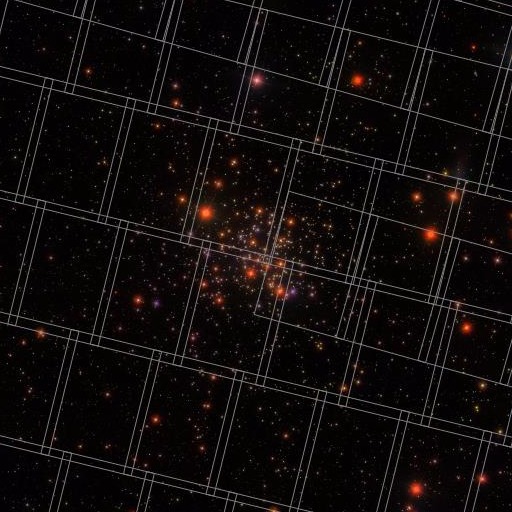}
\caption{Sample data from the Sloan Digital Sky Survey (SDSS). Image boundaries appear as grey lines. All images have the same rectangular size; there is overlap.}
\label{fig:sdss_fields}
\end{figure}

The principal products of astronomical imaging surveys are catalogs of light sources, such as stars and galaxies.
These catalogs are generated by identifying light sources in survey images (e.g., Figure~\ref{fig:sdss_fields}) and characterizing each according to physical parameters such as flux\footnote{Flux is the amount of energy transferred from the light source per unit area (directly facing the light source) per second. ``Apparent brightness'' is another term for flux.}, color, and morphology.

Astronomical catalogs are the starting point for many scientific analyses.
First, catalogs enable demographic inference, which can address fundamental cosmological questions.
For example, researchers may need to know the spatial and luminosity distributions for specific classes (subpopulations) of stars or galaxies.
For these subpopulation-level analyses, accurate quantification of uncertainty in the point estimates of parameters is as important as the accuracy of the point estimates themselves.
It is an open question how to infer a full-sky catalog using Bayesian inference that is also well-calibrated enough for these subpopulation analyses: modeling assumptions that are reasonable for full-sky cataloging are typically too inaccurate for final scientific analysis of subpopulations.
Our work is a step toward creating such a catalog, though this application is not our primary focus.

Second, catalogs inform the design of follow-on surveys using more advanced or specialized instrumentation.
For example, a primary use of the Sloan Digital Sky Survey (SDSS) catalog was to select galaxies to target for follow-up study with a spectrograph~\citep{york2000sloan}.
Whereas image data provides only a rough approximation of the colors of galaxies, spectographs can measure galaxy fluxes for each of hundreds of wavelength bins.
Typically a ``portfolio'' of galaxies to target is selected for each of several galaxy types, e.g., main~\citep{strauss2002spectroscopic}, luminous red galaxies~\citep{eisenstein2001spectroscopic}, and quasars~\citep{richards2002spectroscopic}.
Selecting each portfolio of galaxies amounts to decision making under uncertainty. At present this task is not handled in a statistically coherent way.
Using traditional catalogs, incorporation of uncertainty is not straightforward; astronomers resort to heuristics, typically implemented through cuts based on the raw point estimates appearing in the catalogs. In one case, a portfolio of galaxies was chosen to maximize the sum of the z-values implied by point estimates (and ignoring uncertainties).
In the framework of Bayesian decision theory~\citep{berger2013statistical}, given an approximate posterior distribution, it is straightforward conceptually to select a portfolio of light sources that minimizes a particular cost function.
This second task is our main concern in this work.

\subsection*{Software pipelines for cataloging}

Catalog construction today is based on software pipelines.
For concreteness, we describe the Hyper Suprime-Cam (HSC) software pipeline~\citep{bosch2018hyper}.
The contrasts we subsequently draw between our proposed approach and HSC, however, apply to catalog pipelines in general.
We focus on HSC because it is the state of the art in nearly all respects.
Its code has been merged into the cataloging pipeline for the Large Synoptic Survey Telescope~\citep{lsst}---one of the most important upcoming sky surveys.
HSC draws upon the algorithmic development of both the SDSS software pipeline~\citep{lupton2001sdss}
and SExtractor~\citep{bertin1996sextractor}.

The HSC software pipeline comprises a sequence of steps, including 1) detrending, 2) cosmic ray masking, 3) ``repair'' of saturated and bad pixels through interpolation,
4) estimating the sky background, 5) detecting candidate sources, i.e., localized ``peaks'', 6) estimating the centroids of light sources, 7) estimating the shape of light sources,
8) estimating the flux of light sources, 9) matching light sources to external catalogs,
10) estimating the point-spread function, and 11) performing star/galaxy separation.
Most of these steps depend on estimates from other steps, and many have circular dependencies.
Steps with circular dependencies are repeated multiple times.
For example, at first a circular Gaussian serves as a crude approximation of a star for masking cosmic rays.
Later the cosmic ray detector is rerun with a refined star model.

For the initial sequence of steps (i.e., a ``stage''), the semi-iterative sequence steps are executed on all images independently,
regardless of any overlap.
During later stages, constraints are added that require the algorithm to use a shared estimate for a light source in an overlapping region.
The matching itself depends on aligning the images correctly, which in turn depends on correctly detecting light sources---an additional circular dependency.
Ultimately, aligned, calibrated, and deblended images are ``co-added'' (superimposed) to create one image for each light source.
The final estimate of a light source's properties is based on
the co-added images and accompanying per-pixel variance estimates.

The uncertainty estimates for a light source's flux include only this pixel-level variability.
They do not account for all the other sources of uncertainty that cannot reasonably be modeled as independent across pixels: uncertainty about the light source's centroid, the number of light sources, the image alignments, cosmic ray detection, light sources' shapes, and nearby light sources' fluxes and shapes. The reported uncertainties are based on a Gaussian statistical model of pixels, but one that conditions on the previous stages' estimates of all these quantities.
Effectively, the reported uncertainties are for a conditional distribution rather than a marginal distribution.

Modern cataloging pipelines have struck a balance between algorithmic efficiency and statistical rigor
that has enabled much of the progress in astronomy to date.
Upcoming surveys, however, will probe deeper into the visible universe, creating new challenges.
In particular, whereas blending currently affects just a small number of light sources,
in LSST it is estimated that 68\% of light sources will be blended, requiring new approaches to deblending~\citep{bosch2018hyper}.
In addition, new approaches may let us better interpret existing survey data.
Our aim in this work is to put catalog construction on sounder statistical footing.

\subsection*{Bayesian inference for cataloging}

Our first contribution is a statistical model (Section~\ref{sec:model}) that can simultaneously find centroids,
determine photometry (flux, color, and galaxy morphology), deblend overlapping light sources, perform star/galaxy separation, and adjust estimates of all quantities based on prior information.
Our procedure for all these tasks is based on a single probabilistic model.
The properties of cataloged light sources are modeled as unobserved random variables.
The number of photons recorded by each pixel is modeled by a Poisson distribution with a rate parameter unique to the pixel.
The posterior distribution induced over the unobserved physical properties of the light sources encapsulates knowledge about the catalog's entries, combining prior knowledge of astrophysics with survey imaging data in a statistically efficient manner.
With the model, we can reason about uncertainty
for any quantity in our catalog without conditioning on other estimates being known exactly.

Unfortunately, exact Bayesian posterior inference is intractable for most probabilistic models of interest~\citep{bishop2006pattern}, including this one.
Approximate Bayesian inference is an area of active research. Markov chain Monte Carlo (MCMC) is the most common approach.
Two recent studies demonstrate that Bayesian modeling is the gold standard for astronomical inference, while casting doubt on whether MCMC is viable for constructing a whole astronomical catalog.
~\citet{brewer2013probabilistic} use a single 10,000-pixel image as the dataset for an MCMC procedure.
Obtaining samples from the posterior distribution takes one day using a modern multi-core computer.
\citet{portillo2017improved} run
twelve Intel Xeon cores for an entire day to yield useful results on a similar dataset.
The Sloan Digital Sky Survey---a modern astronomical survey---contains over a billion times as many pixels as these test images.
The upcoming Large Synoptic Survey Telescope (LSST) will collect at least ten terabytes nightly---hundreds of petabytes in total \citep{lsst}. Even basic management of these data requires substantial engineering effort.

Before our work, Tractor~\citep{tractor} was the only program for Bayesian posterior inference that had been applied to a complete modern astronomical imaging survey. Tractor is unpublished work.
It relies on the Laplace approximation: the posterior is approximated by a multivariate Gaussian distribution centered at the mode, having a covariance matrix equal to the negative Hessian of the log-likelihood function at that mode. This approximation is not suitable for either categorical random variables or random variables with multi-modal posteriors---no Gaussian distribution approximates them well. Additionally, because Laplace approximation centers the Gaussian at the mode of the target, rather than the mean, the solution depends on the problem parameterization~\citep{bishop2006pattern}.

Variational inference (VI) is an alternative to MCMC and the Laplace approximation. Like the latter, it uses numerical optimization, not sampling, to find a distribution that approximates the posterior~\citep{blei2017variational}. In practice, the resulting optimization problem is often orders of magnitude faster to solve compared to MCMC approaches. It can be simpler, too. Whereas MCMC transition operators must satisfy strict constraints for validity, the variational optimization problem can in principle be solved using any off-the-shelf technique for numerical optimization. Scaling VI to large datasets is nonetheless challenging.

Our second contribution is to develop and compare two approximate posterior inference procedures for our model: one based on MCMC (Section~\ref{sec:mcmc}) and the other based on VI (Section~\ref{sec:vi}).
Neither is a routine application of Bayesian machinery.
The MCMC procedure combines annealed importance sampling and slice sampling \citep{neal2001annealed, neal2003slice}.
The VI procedure breaks with tradition by optimizing with a variant of Newton's method instead of closed-form coordinate ascent.
For synthetic images drawn from our model, MCMC better quantifies uncertainty, whereas for real astronomical images taken from the Sloan Digital Sky Survey (SDSS), model misspecification may be a more significant limitation than the choice of posterior approximation~(Section~\ref{sec:experiments}).

For either type of data, our VI procedure is orders of magnitude faster than our MCMC procedure.
We scale our VI procedure to the entire Sloan Digital Sky Survey (SDSS) using a supercomputer (Section~\ref{sec:at-scale}).
To our knowledge, this is the largest-scale reported application of VI by at least one order of magnitude.

While our statistical model and inference procedures are accurate on average, the final scientific analysis of a subpopulation of stars or galaxies typically requires priors that are accurate for that particular subpopulation. Several strategies are available for downstream tasks requiring priors specific to the subpopulation, both with and without reprocessing the image data (Section~\ref{sec:discussion}).
\section{Statistical model}
\label{sec:model}
Stars and galaxies radiate photons. An astronomical image records photons---each originating from a particular
celestial body or from background atmospheric and detector noise---that pass through a telescope's lens during an exposure.
A single image contains photons from many light sources; even a single pixel may capture photons from multiple sources.

\begin{figure}[ht!]
	\includegraphics[width=1.8in]{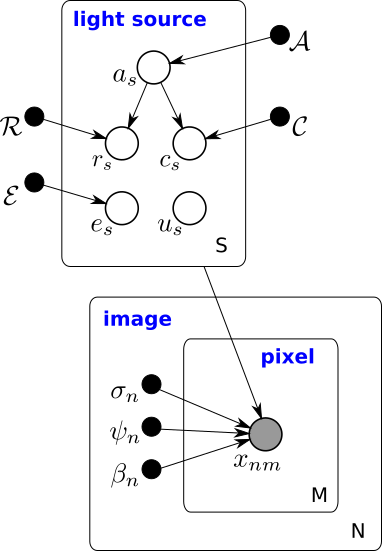}
	\caption{The proposed graphical model. The shaded vertex represents observed
		random variables. Empty vertices represent latent random variables.
		Black dots represent constants, set before inference takes place.
		Edges signify conditional dependencies. Rectangles
		(``plates'') represent independent replication.
		Tables~\ref{structural-table},~\ref{rv-table}, and~\ref{params-table} summarize the variables.
	}
	\label{graphical_model}
\end{figure}

Section~\ref{light-sources} describes our model of light sources. Quantities of interest, such as direction\footnote{A direction is a position on the celestial sphere. We use the term ``direction'', not ``location'', because the distance to a light source, unlike its direction, is not directly observable.}, color, and flux, are random variables. Section~\ref{images} describes a generative model of astronomical images: the distribution of each pixel's intensity---an observed random variable---depends on the latent variables that we aim to infer. Pixel intensities are conditionally independent given these latent random variables.
Figure~\ref{graphical_model} presents our statistical model as a graphical model.

Table~\ref{structural-table} lists the model's structural constants,
denoted by capital Roman letters.
All are positive integers.
None are estimated.

Table~\ref{rv-table} lists the model's random variables for a light source $s \in \{1,\ldots,S\}$,
an image $n \in \{1,\ldots, N\}$,
and a pixel $m \in \{1,\ldots,M\}$.
(S, N, and M appear in Table~\ref{structural-table}.)
All are denoted by lowercase Roman letters.
All are scalars except for the color vector $c_s$ and the direction vector $u_s$.
Inferring the posterior distribution of the unobserved random variables in Table~\ref{rv-table} is the primary problem addressed by this article.

Table~\ref{params-table} lists model parameters.
The first eight rows describe hyperparameters; they parameterize the prior and are distinguished by calligraphic font.
They are estimated a priori by maximum likelihood, as described in Section~\ref{prior}.
The remaining parameters, denoted by lowercase Greek letters, are set by the SDSS pipeline.

\subsection{Light sources}
\label{light-sources}

An astronomical catalog is a table with one row for each light source.
The number of light sources, $S$, is treated as a known constant here; we determine it by running existing cataloging software~\citep{bertin1996sextractor}.
Modeling $S$ as random we defer to future work.

Light sources in our model are either stars or galaxies, as are the vast majority of light sources in the universe.
Exceptions include asteroids, planets, airplanes, and man-made satellites, which also occasionally appear in astronomical images.
For light source $s=1,\ldots,S$, the latent random variable
\begin{align}
a_s &\sim \mathrm{Bernoulli}(\mathcal A)
\end{align}
indicates whether it is a star (${a_s=1}$) or a galaxy (${a_s=0}$). Here $\mathcal{A}$ is the prior probability that a light source is a star. (We discuss how we set $\mathcal A$, and all other prior hyperparameters, in Section~\ref{prior}.)

\FloatBarrier
\begin{table}
\centering
\caption{Structural constants in our model.}
\label{params}
\centering
\begin{tabular}{lll}
\toprule
\textbf{name} & \textbf{brief description}  & \textbf{SDSS value} \\
\hline
B & number of filter bands & 5\\
E & number of PSF ``eigenimages'' & 4\\
F & number of knots per PSF eigenimage & $51 \times 51$\\
H & number of rows of pixels per image & 2048 \\
I & number of source types (i.e., star, galaxy) & 2\\
J & number of components in the color prior mixture & 8\\
K & number of components in the galaxy mixture model & 8\\
L & number of parameters in a WCS header & 16\\
M & number of pixels per image & $H \times W$\\
N & number of images & 4,690,230\\
Q & number of knots for the sky background model & $192 \times 256$\\
S & number of light sources & 469,053,874\\
W & number of columns of pixels per image & 1361 \\
\hline
\end{tabular}
\label{structural-table}
\end{table}

\begin{table}
\centering
\caption{Random variables in our model.}
\centering
\begin{tabular}{llll}
\hline
\textbf{name} & \textbf{brief description} & \textbf{units} & \textbf{domain} \\
\hline
$a_s$           & galaxy / star indicator & unitless & \{0, 1\} \\
$c_s$           & colors & magnitude & $\mathbb R^{B-1}$ \\
$e_s^{angle}$   & angle of galaxy's major axis & degrees & $[0, 180)$ \\
$e_s^{radius}$  & galaxy's half-light radius & arcseconds & $(0, \infty)$ \\
$e_s^{profile}$ & galaxy's profile mixing weight & unitless & $[0, 1]$ \\
$e_s^{axis}$    & galaxy's minor-major axis ratio & unitless & $(0, 1)$ \\
$r_s$           & reference-band flux density & nanomaggies & $[0, \infty)$ \\
$u_s$           & direction (longitude, latitude) & degrees & $[0, 360) \times [-90, 90]$ \\
$x_{nm}$        & pixel intensity (observed) & photon count & $\{0,1,2,\ldots\}$ \\
\hline
\end{tabular}
\label{rv-table}
\end{table}

\begin{table}
\centering
\caption{Parameters in our model.}
\centering
\begin{tabular}{lll}
\hline
\textbf{name} & \textbf{brief description}  & \textbf{domain} \\
\hline
$\mathcal A$           & prior probability a light source is a star & $[0, 1]$\\
$\mathcal C^{weight}$  & color prior mixture weights &  $\mathbb R^{I \times J}$\\
$\mathcal C^{mean}$    & color prior mixture component means &  $\mathbb R^{I \times J \times (B - 1)}$\\
$\mathcal C^{cov}$     & color prior mixture component covariance matrices &  $\mathbb R^{I \times J \times (B - 1)  \times (B - 1)}$\\
$\mathcal E^{radius}$  & galaxy half-light radius prior parameters  & $\mathbb R^2$\\
$\mathcal E^{profile}$ & galaxy profile prior parameters & $\mathbb R^2$\\
$\mathcal E^{axis}$    & galaxy axis ratio prior parameters & $\mathbb R^2$\\
$\mathcal R$           & reference-band flux prior parameters & $\mathbb R^{I \times 2}$\\
\hline
$\sigma_{n}$           & sky background model & $\mathbb R^Q$\\
$\psi_n^{calib}$       & expected number of photons per nanomaggy & $\mathbb R^H$\\
$\psi_n^{wcs}$         & image alignment & $\mathbb R^L$\\
$\psi_n^{weight}$      & point spread function loadings & $\mathbb R^E$\\
$\psi_n^{image}$       & point spread function principal components & $\mathbb R^{E \times F}$\\
$\beta_n $             & filter band & $\{1,2,\ldots,B\}$\\
\hline
\end{tabular}
\label{params-table}
\end{table}
\FloatBarrier

The latent random two-vector $u_s$ denotes the direction of light source $s$
in the units of J2000.0 equatorial coordinates, a latitude and longitude system relative to the Earth's equator.
Figure~\ref{sdss_coverage} illustrates this system of coordinates.
The first coordinate is longitude and the second coordinate is latitude.
Both are measured in degrees.

A priori, $u_s$ is uniformly distributed over the sphere.
Treating light sources as uniformly distributed is a simplification---some regions of the sky are known a priori to have more light sources than others, e.g., the galactic plane. This is known as directional dependence. Additionally, it is a simplification to model light sources as positioned independently of each other; gravity causes some clustering among light sources.

\subsubsection{Flux}
\label{flux}

The flux of light source $s$ is defined as its expected total radiation reaching a unit area of Earth's surface directly facing $s$, per unit of time. We can measure the flux as the portion of this radiation (per square meter per second) that passes through each filter in a standardized filter set. Such a set is called a filter system. These standardized filters are approximately band-pass: each allows most of the energy in a certain band of wavelengths through, while blocking most of the energy outside the band.
The physical filters attached to a telescope lens closely match the standardized filters of some filter systems.

The five SDSS filters are named for the wavelengths they are most likely to let pass:
ultraviolet~($u'$), green~($g'$), red~($r'$), near infrared~($i'$), and infrared~($z'$).
Figure~\ref{sdss-filter-curves} shows how likely a photon of a particular wavelength is to pass through each filter.
\cite{fukugita1996sloan} further describe the SDSS filter system.

We model flux with respect to the $B = 5$ filters of the SDSS filter system.
We designate a particular filter as the ``reference'' filter, letting the
random variable $r_s$ denote the flux of object $s$ with respect to that
filter. A priori,
\begin{align}
r_s | (a_s = i) &\sim \mathrm{LogNormal}(\mathcal R_{i1}, \mathcal R_{i2}),\,\,\, i \in \{0,1\}.
\end{align}
Our prior depends on $a_s$ to reflect that stars tend to have higher flux density than galaxies.
The flux density $r_s$ is measured in nanomaggies~\citep{sdssglossary,nanomaggies}.
One nanomaggy is equivalent to $3.631 \times 10^{-6}$ Jansky.
A nanomaggy is a linear unit; we expect to receive twice as many photons from a two-nanomaggy light source as from a one-nanomaggy light source.

The log-normal distribution reflects that flux is non-negative and that stars' fluxes often differ by orders of magnitude. Empirically,
a log-normal distribution fits the SDSS catalog better than any gamma distribution---another common model for non-negative real-valued variables.
In future work, we may also explore a power law distribution for galaxy fluxes, as there is some theoretical support for that model.

The flux of light source $s$ with respect to the remaining $B-1$ filters is encoded using colors.
The color $c_{s\beta}$ is defined as the log ratio of
fluxes with respect to filters $\beta$ and $\beta + 1$. Here, the filters are
ordered by the wavelength bands they let pass. The $B-1$ colors for
object $s$ are collectively denoted by $c_s$, a random $(B-1)$-vector.
We denote the colors as \textit{u-g, g-r, r-i}, and \textit{i-z}.
The reference-filter flux $r_s$ and the colors $c_s$ uniquely specify the flux for light source $s$ through any filter $\beta$, denoted $\ell_{s\beta}$.

\begin{figure}
  \includegraphics[width=3.5in]{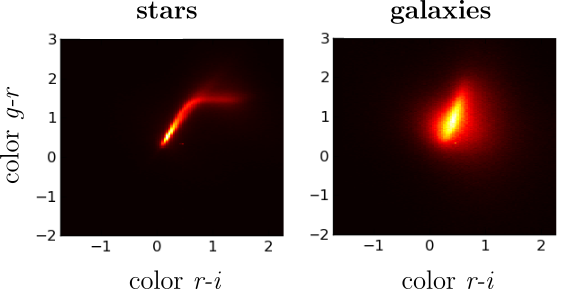}
  \caption{Density plots for two colors, \textit{g-r} and \textit{r-i}, based on the SDSS DR10 catalog.}
  \label{color_priors}
\end{figure}

Our model uses the color parameterization because stars and galaxies have very
distinct prior distributions in color space. Indeed, for idealized
stars---blackbodies---all $B - 1$ colors lie on a one-dimensional manifold
indexed by surface temperature. On the other hand, though galaxies are
composed of stars, theory does not suggest they lie near the same
manifold: the stars in a galaxy can have many different surface
temperatures, and some of the photons are re-processed to other energies
through interactions with dust and gas.
Figure~\ref{color_priors} demonstrates that stars are much closer to a
one-dimensional manifold in color space than galaxies are.

We model the prior distribution on $c_s$ as a $D$-component Gaussian mixture model (GMM):
\begin{align}
c_s | (a_s = i) &\sim \mathrm{GMM}(\mathcal C_{i}^{weight}, C_{i}^{mean}, \mathcal C_{i}^{cov}),\,\,\, i \in \{0,1\}.
\end{align}
We discuss how we set $D$ and the color priors' hyperparameters in Section~\ref{prior}.

\subsubsection{Spatial extent}

\begin{figure}
\begin{floatrow}
\ffigbox{%
  \includegraphics[width=1.8in]{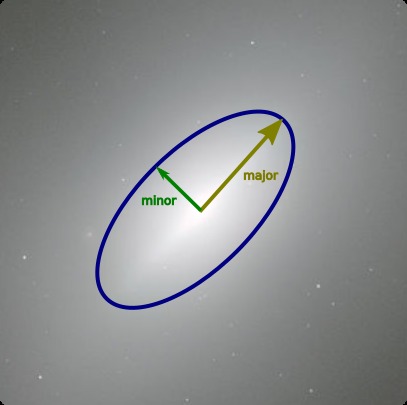}
}{%
  \caption{A schematic of the galaxy light kernel. The blue ellipse surrounds half of the light emissions of this galaxy.
  The length of the major axis is the half-light radius $e_s^{radius}$. The angle in degrees of the major axis is~${e_s^{angle} = 45}$. The ratio of the lengths of minor and major axes is~${e_s^{axis} = 1/2}$. Because this galaxy is purely elliptical,~${e_s^{profile} = 0}$.}%
  \label{cartoon-galaxy}
}
\ffigbox{%
  \includegraphics[width=1.6in]{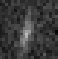}
}{%
  \caption{A distant galaxy approximately 20 pixels
  in height, estimated to
  have half-light radius~${e_s^{radius}=0.6}$ arcseconds, rotation angle~${e_s^{angle}=80}$ degrees, and minor-major axis ratio~${e_s^{axis}=0.17}$.}
  \label{fig:faint_galaxy}
}
\end{floatrow}
\end{figure}

\begin{figure}
\begin{subfigure}{0.45\textwidth}
\includegraphics[height=1.8in]{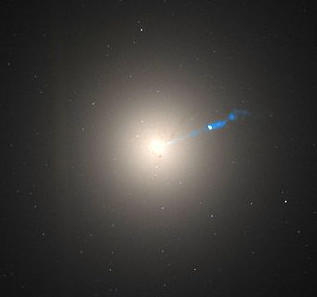}
\subcaption{Messier 87, a galaxy that exhibits the de Vaucouleurs profile. Credit: NASA}
\label{fig:dev_galaxy}
\end{subfigure}
\hfill
\begin{subfigure}{0.45\textwidth}
\includegraphics[height=1.8in]{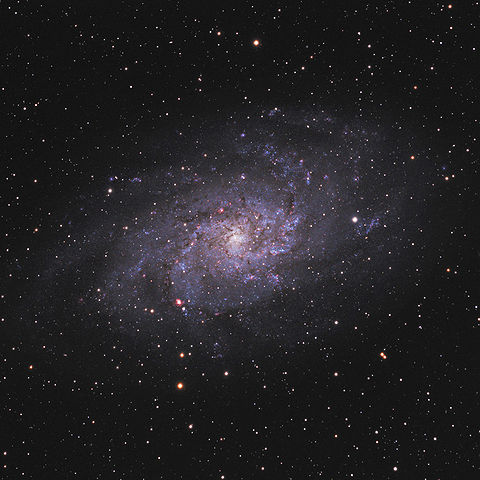}
\subcaption{Triangulum, a galaxy that exhibits the exponential profile. Credit: NASA}
\label{fig:exp_galaxy}
\end{subfigure}
\caption{Extremal galaxy profiles.}
\label{galaxy-profiles}
\end{figure}

Consider a light source~$s$, centered at some direction~$u_s$.
Its flux density in filter band~$\beta$, measured at a possibly different direction~$\mu$,
is given by
\begin{align}
\varphi_{s\beta}(\mu) \coloneqq h_s(\mu) \ell_{s\beta}.
\end{align}
Here $h_s$ (a density) models the spatial characteristics of light source $s$, quantifying its relative intensity at each direction $\mu$ specified in sky coordinates (not image-specific ``pixel coordinates''). We refer to $h_s$ as the ``light kernel'' for light source $s$.

The distance from Earth to any star other than the Sun exceeds the star's radius by many orders of magnitude. Therefore, we model stars as point sources. If light source $s$ is a star (i.e., $a_s = 1$), then $h_s$ is simply a delta function: one if $\mu = u_s$, zero otherwise.

Modeling the two-dimensional appearance of galaxies as seen from Earth is more involved. If light source $s$ is a galaxy (i.e., ${a_s = 0}$), then $h_s$ is parameterized by a latent random 4-vector
\begin{align}
e_s \coloneqq (e_s^{profile}, e_s^{angle}, e_s^{radius}, e_s^{axis}).
\end{align}
We take $h_s$ to be a convex combination of two extremal profiles,
known in astronomy as ``de Vaucouleurs'' and ``exponential''
profiles:
\begin{align}
h_{s}(\mu) &= e_s^{profile} h_{s1}(\mu) + (1 - e_s^{profile}) h_{s2}(\mu).
\end{align}
The de Vaucouleurs profile is characteristic of elliptical galaxies, whose luminosities vary gradually in space (Figure~\ref{fig:dev_galaxy}), whereas the exponential profile matches spiral galaxies (Figure~\ref{fig:exp_galaxy})~\citep{feigelson2012modern}.
The profile functions $h_{s1}(\mu)$ and $h_{s2}(\mu)$ also account for additional galaxy-specific parameters illustrated in Figure~\ref{cartoon-galaxy}. In particular,
each profile function is a rotated, scaled mixture of bivariate normal distributions. Rotation angle and scale are galaxy-specific, while the remaining parameters of each mixture are not:
\begin{align}
  h_{si}(\mu)
  & = \sum_{j=1}^{J} \alpha_{ij} \phi(\mu; u_{s}, \tau_{ij} \Sigma_{s}),\,\,\, i \in \{0,1\}.\label{eq:hsi}
\end{align}
Here the $\alpha_{ij}$ and the $\tau_{ij}$ are prespecified constants that characterize the exponential and de Vaucouleurs profiles; $u_s$ is the center of the galaxy in sky coordinates; $\Sigma_s$ is a 2$\times$2-covariance matrix shared across the components; and $\phi$ is the bivariate normal density.

The light kernel $h_s(\mu)$ is a finite scale mixture of Gaussians: its mixture
components have a common mean $u_s$; the isophotes (level sets of
$h_s(\mu)$) are concentric ellipses. Although this model
prevents us from fitting individual ``arms,'' like those of the galaxy in
Figure~\ref{fig:exp_galaxy}, most galaxies
are not sufficiently resolved to see such substructures.
Figure~\ref{fig:faint_galaxy} shows a more typical galaxy image.

The spatial covariance matrix~$\Sigma_s$ is parameterized by a rotation angle~$e_s^{angle}$, an eccentricity (minor-major axis ratio)~$e_{s}^{axis}$, and an
overall size scale~$e_{s}^{radius}$:
\begin{align}
\Sigma_s \coloneqq R_s^{\top}\begin{bmatrix} [ e_{s}^{radius}]^{2} & 0\\
                            0 & [e_{s}^{axis}]^2[e_{s}^{radius}]^{2}
\end{bmatrix} R_s,
\end{align}
where the rotation matrix is given by
\begin{align}
R_s \coloneqq \begin{bmatrix}\cos e_s^{angle} & -\sin e_s^{angle}\\
                           \sin e_s^{angle} & \cos e_s^{angle}
\end{bmatrix}.
\end{align}
The scale $e_{s}^{radius}$ is specified in terms of half-light radius---the
radius of the disc that contains half of the galaxy's light emissions
before applying the eccentricity $e_s^{angle}$.

All four entries of $e_s$ are random.
The mixing weight prior is given by
\begin{align}
e_s^{profile} &\sim \mathrm{Beta}(\mathcal E^{profile}_1, \mathcal E^{profile}_2).
\end{align}
Every angle is equally likely, and galaxies are symmetric, so
\begin{align}
e_s^{angle} &\sim \mathrm{Uniform}([0, 180]).
\end{align}
We found that the following half-light-radius distribution fit well empirically:
\begin{align}
e_s^{radius} &\sim \mathrm{LogNormal}(\mathcal E^{radius}_1, \mathcal E^{radius}_2).
\end{align}
The ``fatter'' tail of a log-normal distribution fits better than a gamma distribution, for example.
A priori, the minor-major axis ratio is beta distributed:
\begin{align}
e_s^{axis} &\sim \mathrm{Beta}(\mathcal E^{axis}_1, \mathcal E^{axis}_2).
\end{align}

\subsubsection{Setting the priors' parameters}
\label{prior}
The light source prior is parameterized by 1099 real-valued scalars.
All but ten are for the GMM color prior.
Empirical Bayes is an appealing way to fit this prior because the number of parameters is small relative to the number of light sources (hundreds of millions for SDSS).

Unfortunately, re-fitting the prior parameters iteratively during inference---a common way of performing empirical Bayes---is difficult in a distributed setting: fitting the global prior parameters during inference couples together numerical optimization for disparate regions of sky.
Instead, we fit the prior parameters based on existing SDSS catalogs through maximum likelihood estimation.
Because these prior parameters are fit to a catalog based on the same data we subsequently analyze, our procedure is in the spirit of empirical Bayes.
However, using maximum likelihood in this way to assign priors ignores measurement error (and classification error) and therefore will produce priors that are  overdispersed. It produces estimates that are formally inconsistent, unlike conventional empirical Bayes approaches that iteratively refit the prior.

If the depth of our catalog were much greater than existing SDSS catalogs, we too might refit these
prior parameters periodically while performing inference.
Refitting in this way could be interpreted as a block coordinate ascent scheme.
However, in our work to date, the depth of our catalog is limited by the peak-finding preprocessing routine, just as in SDSS.
Therefore, for simplicity, we hold these prior parameters fixed during inference.

Fitting the color prior warrants some additional discussion.
First, maximum-likelihood estimation for a GMM is nonconvex, so the optimization path may matter: we use the GaussianMixture.jl software~\citep{gmm}.
Second, we set the number of GMM components $D$ based on computational considerations.
In principle, $D$ could be set with a statistical model-selection criterion.
In practice, we set $D=8$ without any apparent accuracy reduction for the point estimates, which is the primary way we assess our model in Section~\ref{sec:experiments}.
Because we have so much data (millions of light sources), there is no risk of overfitting with $D=8$: held-out log-likelihood improves as $D$ increases up to $D=256$, the largest setting our hardware allowed us to test.
There is also little risk that $D=8$ underfits: setting $D=16$ does not substantively change our estimates.

Empirical Bayes seems broadly applicable to sky-survey data; the number of light sources in typical surveys is large relative to the number of hyperparameters.
But the details of our procedure (e.g., how to set $D$, or whether to update the hyperparameters iteratively during inference) may need to be tailored based on the research goals. If so, our fitted priors may be considered ``interim'' priors.

\subsection{Images}
\label{images}

\begin{figure}
\begin{floatrow}
\ffigbox{%
\includegraphics[width=.6\linewidth]{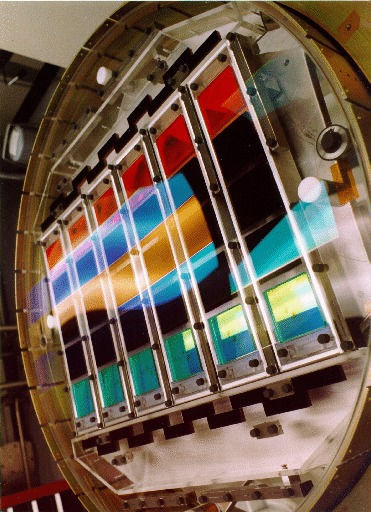}
}{%
\caption{The SDSS camera. Its CCDs---each $2048 \times 2048$ pixels---are arranged in six columns and five rows. A different filter covers each row. Credit:~\cite{sdsscamera}.}
\label{sdss-camera}
}
\ffigbox{%
\includegraphics[width=\linewidth]{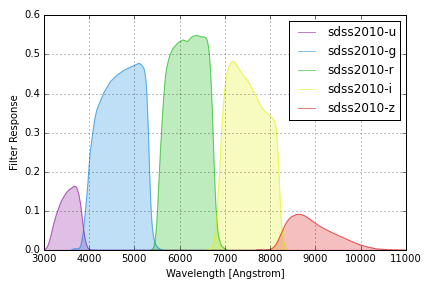}
}{%
\caption{SDSS filter curves. Filter response is the probability that a photon of a particular wavelength will pass through the filter. Credit:~\cite{doi2010photometric}.}
\label{sdss-filter-curves}
}
\end{floatrow}
\end{figure}

\begin{figure}[t]
\includegraphics[width=3in]{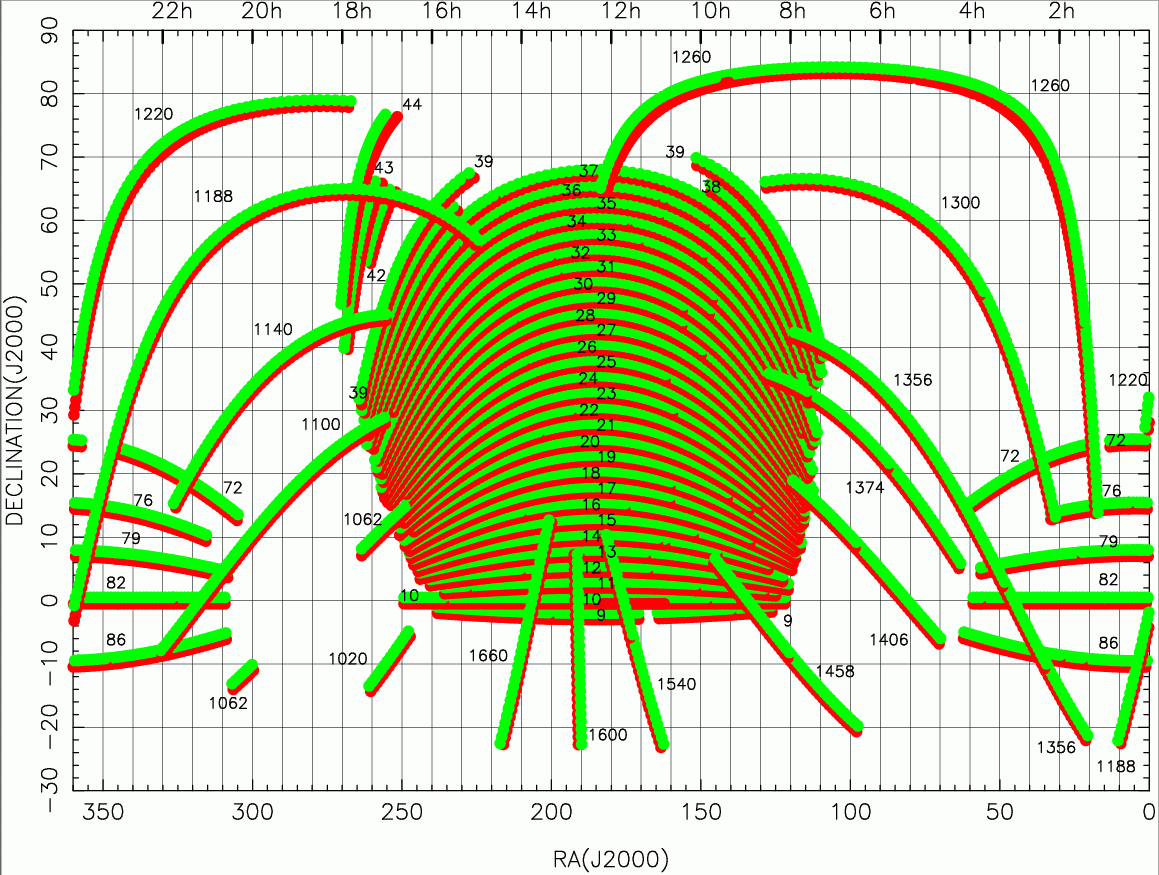}
\caption{SDSS sky coverage map. Each monocolor arc represents the sky
photographed during a particular night. Axes' units are degrees of right ascension (longitude) and declination (latitude). Credit:~\cite{sdsscoverage}.}
\label{sdss_coverage}
\end{figure}

Astronomical images are taken through telescopes.
Photons that enter the telescope reach a camera that records the pixel each photon hits,
thus contributing an electron.
The SDSS camera (Figure~\ref{sdss-camera}) consists of 30 charge-coupled devices (CCDs) arranged in a grid of six columns and five rows.
Each row is covered by a different filter---transparent colored glass that limits which photons can pass through and potentially be recorded.
Each of the five filters selects, stochastically, for photons of different wavelengths (Figure~\ref{sdss-filter-curves}).
Multiple images of the same region of the sky with different filters reveal the colors of stars and galaxies.

The SDSS telescope collects images by drift scanning, an imaging regime where the camera reads the CCDs continuously as the photons arrive.
Each night the telescope images a contiguous ``arc'' of sky (Figure~\ref{sdss_coverage}).

Each arc is divided into multiple image files.
SDSS Data Release 13 contains $N=$ 4,690,230 of these images, each taken through one of the 30 CCDs.
For $n=1,\ldots,N$, the constant $\beta_n$ denotes the filter color for image $n$.

Each image is a grid of $M=2048 \times 1361$ pixels.
The random variable $x_{nm}$ denotes the count of photons that, during the exposure for image $n$, entered the telescope, passed through the filter, and were recorded by pixel $m$.

\subsubsection{Skyglow}

The night sky is not completely dark, even in directions without resolvable light sources. This is due to both artificial light production (e.g., light pollution from cities) and natural phenomena.
The background flux is called ``skyglow.''
Sources of natural skyglow include sunlight reflected off dust particles in the solar system, nebulosity (i.e., glowing gas---a constituent of the interstellar medium), extragalactic background light from distant unresolved galaxies, night airglow from molecules in Earth's atmosphere, and scattered starlight and moonlight.
The flux from skyglow (``sky intensity'') varies by the time of the exposure, due to changing atmospheric conditions.
It also varies with direction; for example, sky intensity is typically greater near the galactic plane.
We model skyglow as a spatial Poisson process whose rate varies gradually by pixel, independent of stars and galaxies.
For the vast majority of pixels, the skyglow is the only source of photons.

Sky intensity is estimated during preprocessing by pre-existing software \citep{bertin1996sextractor} and fixed during inference.
This software fits a smooth parametric model to the intensities of the pixels that it determines are not near any light source.
The sky intensity could, in principle, be fit within our inference procedure; we defer this idea to future work.

The sky intensity for image $n$ is stored as a grid of $Q$ intensities in the matrix $\sigma_n$.
Typically $Q \ll M$ because the sky intensity varies slowly.
To form the sky intensity for a particular pixel, $\sigma_n$ is interpolated linearly.
We denote the sky intensity for a particular pixel by $\sigma_{n}(m)$.

\subsubsection{Point-spread functions}

Astronomical images are blurred by a combination of small-angle scattering in Earth's atmosphere, the diffraction limit of the telescope, optical distortions in the camera, and charge diffusion within the silicon of the CCD detectors. Together these effects are represented by the ``point-spread function'' (PSF) of a given image. Stars are essentially point sources, but the PSF represents how their photons are spread over dozens of adjacent pixels.

The PSF is set during preprocessing by pre-existing software~\citep{lupton2001sdss}. This software fits the PSF based on several stars with extremely high flux in each image whose characteristics are well established by previous studies using different instrumentation (e.g., spectrographs).
As with sky intensity, we could fit the PSF jointly with light sources through our approximate inference procedure, but we do not pursue this idea here.

The PSF is specified through several image-specific parameters that are bundled together in $\psi_n$.
The vector $\psi_n^{calib}$ gives the expected number of photons per nanomaggy for each column of image $n$.
The vector $\psi_n^{wcs}$ specifies a mapping from sky direction to pixel coordinates. This mapping is linear---an approximation that holds up well locally.
The rows of the matrix~$\psi_n^{image}$ give the top principal components from an eigendecomposition of the PSF as rendered on a grid centered at a light source.
The vector~$\psi_n^{weight}$ gives the loading of the PSF at any point in the image. It has smooth spatial variation.

Consider a one-nanomaggy star having direction $\mu$.
We denote its expected contribution of photons to the $m$th pixel of image~$n$ as~$g_{nm}(\mu)$; this is derived as needed from the explicitly represented quantities discussed above.

\subsubsection{The likelihood}

Let $z_s \coloneqq (a_s, r_s, c_s, e_s, u_s)$ denote the latent random variables for light source $s$.
Let $z \coloneqq \{z_s\}_{s=1}^S$ denote all the latent random variables.
Then, for the number of photons received by pixel $m$ of image $n$, we take the likelihood to be
\begin{align}
x_{nm} | z \sim \mathrm{Poisson}(\lambda_{nm}).
\label{xnm}
\end{align}

The dependence of $\lambda_{nm}$ on $z$ is not notated here.
We model $x_{nm}$ as observed, though the reality is more complicated~\citep{frame}:
at the end of an exposure, the CCD readout process transfers the electrons to a small capacitor, converting the (discrete) charge to a voltage that is amplified and forms the output of the CCD chip.
The net voltage is measured and digitized by an analog-to-digital converter (ADC).
The conversion is characterized by a conversion gain.
The ADC output is an integer called a digital number (DN).
The conversion gain is specified in terms of electrons per DN.
While in our model $x_{nm}$ is the number of photons received, in practice we set $x_{nm}$ to a value determined by scaling DN according to gain and rounding it to the nearest integer.
The Poisson mass function is fairly constant across the quantization range.
\footnote{An alternate perspective is that $x_{nm}$ is not approximated in practice: rather, $x_{nm}$ is our approximation.
We do not take this perspective so that we can explain our model at a high level before introducing low-level details of CCD technology.
}

Because of these complexities, it is not clear whether a Poisson distribution is more suitable here than a Gaussian distribution with its mean equal to its variance.
We make no claims about the superiority of one or the other.
In the SDSS, the sky background is typically at least 500 electrons per pixel, so it seems unlikely that the choice of a Gaussian (with its mean equal to its variance) or a Poisson distribution would matter.
Furthermore, neither likelihood simplifies the subsequent inferential calculations.

In Equation~\ref{xnm}, the rate parameter $\lambda_{nm}$ is unique to pixel $m$ in image $n$.
It is a deterministic function of the catalog (which includes random quantities)
given by
\begin{align}
\lambda_{nm} &\coloneqq \sigma_{n}(m) + \sum_{s=1}^S \ell_{s\beta_n} \int g_{nm}(\mu) h_s(\mu) \,d\mu\,.
\end{align}
The summation over light sources reflects the assumption that light sources do not occlude one another, or the skyglow.
The integral is over all sky locations.
In practice, it can be restricted to pixels near pixel $m$---distant light sources contribute a negligible number of photons.
Our implementation bases this distance measurement on conservative estimates of light sources' extents from SExtractor~\citep{bertin1996sextractor}.
As shorthand, we denote the integral as
\begin{align}
f_{nms} \coloneqq \int g_{nm}(\mu) h_s(\mu)\,d\mu.
\end{align}
If light source $s$ is a star, then it is straightforward to express $f_{nms}$ analytically:
\begin{align}
f_{nms} = g_{nm}(u_s).
\end{align}

If light source $s$ is a galaxy, the same integral is more complex because galaxies have spatial extent.
Our approach is to approximate $g_{nm}$ with a mixure of bivariate normal densities.
Because Gaussian-Gaussian convolution is analytic, we get an analytic approximation to $f_{nms}$.

Our primary use for the model is
computing the posterior distribution of its unobserved random variables conditional on a
particular collection of astronomical images. We denote the posterior by $p(z | x)$, where
$x \coloneqq \{x_{nm}\}_{n=1,m=1}^{N,M}$ represents all the pixel intensities.
Exact posterior inference is computationally intractable for
the proposed model, as it is for most non-trivial probabilistic models.
The next two sections consider two approaches to approximate posterior inference: Markov chain Monte Carlo (MCMC) and variational inference (VI).

\section{Markov chain Monte Carlo}
\label{sec:mcmc}

Markov chain Monte Carlo (MCMC) is a common approach for approximating posterior distributions in computationally challenging Bayesian models.
MCMC draws samples from a stochastic process on the parameter space whose stationary distribution is the posterior distribution of interest.
The stochastic process is specified by a transition kernel, denoted $\mathcal T$.
The empirical distribution of these samples approximates the posterior distribution.
Statistics of this empirical distribution, such as its mean and its quantiles, approximate the same
statistics of the posterior distribution.

Our problem presents two challenges for MCMC.
First, the state space is extremely high-dimensional---there are multiple random variables for each of millions of light sources.
We cannot consider transition kernels that require hand-tuning of dimension-specific parameters, such as step size, proposal variance, or temperature schedule.
Second, the state space is trans-dimensional. 
Galaxies have more parameters than stars, and light-source types (star/galaxy) are themselves random.

We propose a multi-level sampling procedure.
In an outer loop based on (block) Gibbs sampling \citep{robert2013monte}, 
light sources are treated sequentially. Each light source's latent variables are sampled with any overlapping light sources' latent variables, denoted $z_{-s}$, held fixed.
Formally, in Gibbs iteration $k=1,\ldots,K$, we draw
\begin{align}
z_s^{(k)} \sim p(z_s | x, z_{-s}^{(k-1)}) 
\label{eqzs}
\end{align}
for light sources $s=1,\ldots,S$ in sequence.
To speed up convergence,
we initialize $z_1^{(0)},\ldots,z_S^{(0)}$ with approximately correct values determined by a preprocessing routine.

Our strategy for generating samples from the distribution in Equation~\ref{eqzs} is to first draw a sample from the marginal posterior over the source's type~$a_s$ (star or galaxy) and then draw samples from the conditional posterior over the remaining source parameters, $w_s \triangleq (r_s, c_s, e_s, u_s)$:
\begin{align}
    a_s &\sim p(a_s | x, z_{-s}) && \text{ marginal source type; } \label{eq:source-type-marginal} \\
    w_s ~\vert~ a_s &\sim  p(w_s | a_s, x, z_{-s})) && \text{ conditional source parameters.} \label{eq:source-param-conditional}
\end{align}
To generate a sample from Equation~\ref{eq:source-type-marginal} we use annealed importance sampling (AIS) \citep{neal2001annealed}, initialized with outputs of the AIS step.
To generate a conditional sample from Equation~\ref{eq:source-param-conditional} we use slice sampling~\citep{neal2003slice}. 
We will explain each sampler in turn in Sections~\ref{mcmc1} and \ref{mcmc2}.

Recall that $a_s$ is the Bernoulli random variable that indexes the source type (star/galaxy), and thus the dimension of our state space.
This two-step sampling strategy allows us to avoid using a trans-dimensional sampler like reversible-jump MCMC \citep{green1995reversible}, a technique that requires constructing a potentially complex trans-dimensional proposal function~\citep{fan2011reversible}.

\subsection{Sampling the posterior over $a_s$}
\label{mcmc1}
To generate a sample from the marginal posterior over $a_s$, we estimate the marginal posterior probabilities of $a_s=1$ and $0$ (which together sum to one).
By Bayes's rule, we can write the marginal posterior
\begin{align}
p(a_s = 1 | x, z_{-s}) &\propto p(x | a_s = 1, z_{-s}) p(a_s=1 | z_{-s}) \label{eq:source-type-marginal-bayes}
\end{align}
The term $p(x | a_s=1, z_{-s})$ is the marginal likelihood of the observation $x$ given the source is of type $a_s=1$, which is the type of estimand AIS is designed to estimate.  The term $p(a_s=1 | z_{-s}) = p(a_s=1)$ is the prior over source type.

AIS is an iterative procedure to estimate the normalizing constant (i.e., the integral) of an unnormalized probability density $\pi$.  In order to estimate the marginal likelihood $p(x | a_s, z_{-s})$, we estimate the normalizing constant of the distribution 
\begin{align}
\pi(w_s) \coloneqq p(x | w_s, a_s, z_{-s}) p(w_s | a_s, z_{-s})
\end{align}
for both source types, $a_s = 0$ and $a_s=1$.  
This normalizing constant is $p(x | a_s, z_{-s})$.
Given an estimate of $p(x | a_s, z_{-s})$ (for both settings of $a_s$) and a prior over $a_s$, we can construct an estimate of $p(a_s = 1 | x, z_{-s})$ using Bayes' rule as in Equation~\ref{eq:source-type-marginal-bayes}.
Then we can sample from $p(a_s | x, z_{-s})$.

In addition to the target $\pi$, AIS takes as input a sequence of $T$ distributions
$\pi_0, \pi_1, \dots, \pi_T$ that approach the target.
The statistical efficiency of AIS depends on the similarity of intermediate distributions $\pi_{t-1}(z_s) / \pi_{t}(z_s)$.
We set $\pi_0(z_s) \coloneqq p(w_s | a_s, z_{-s})$---a normalized density.
For $t=1,\ldots,T$, we set
\begin{align}
  \pi_t(w_s) &= \pi_0(w_s)^{1-\gamma_t} \pi(w_s)^{\gamma_t}
\end{align}
for a sequence of temperatures $0 = \gamma_0 < \gamma_1 < \dots < \gamma_T = 1$.
These (unnormalized) distributions interpolate between the prior and the posterior.

For $t=1, \dots, T$, let $\mathcal T_t$ be a Markov chain transition that leaves (the normalized version of) $\pi_t$ invariant.
To implement each transition kernel, $\mathcal{T}_t$, we use slice sampling, a Markov chain Monte Carlo method that requires little tuning and automatically adapts to the local scale for each variable \citep{neal2003slice}.
We iterate over each variable in $z_s$, forming a slice-sampling-within-Gibbs transition kernel.

We begin by sampling $w_s^{(0)} \sim \pi_0$.
Then, for $t=1, \dots, T$, we draw
\begin{align}
w_s^{(t)} | w_s^{(t-1)} \sim \mathcal{T}_t(w_s^{(t-1)}, w_s^{(t)}).
\end{align}
After $T$ iterations, $w_s^{(T)}$ is approximately distributed according to (the normalized version of) $\pi_T = \pi$, and
\begin{align}
	\mathcal Z_s \coloneqq \exp \sum_{t=1}^T \log \frac{\pi_{t}(w_s^{(t-1)})} { \pi_{t-1}(w_s^{(t-1)}) }
\end{align}
is a consistent estimator of $p(x | a_s, z_{-s})$ \citep{neal2001annealed}.
AIS can be viewed as importance sampling over an augmented state space where the expanded dimensions begin with the prior distribution and gradually anneal to the targeted posterior according to $T$ temperatures.
Thus, the ratio of these weights is a consistent estimator of the marginal likelihood.

Estimating the marginal likelihood (also referred to as the model evidence) is a rich area of methodological development. \cite{skilling2004nested} presents another popular approach for computing marginal likelihood estimates, known as nested sampling.  However, \cite{friel2012estimating} show cases where nested sampling is less efficient statistically and computationally than AIS, motivating our use of AIS in this work.

\subsection{Sampling source parameters conditioned on $a_s$}
\label{mcmc2}
The final step of our AIS procedure draws samples from $p(w_s | a_s, x, z_{-s})$.
For each source type (star/galaxy), we run $N'$ independent repetitions of our AIS procedure.
We use the resulting samples as independent starting positions for $N'$ Markov chains.
We run these $N'$ chains for $B'$ more steps, monitoring convergence and mixing criteria \citep{gelman1992inference}.
This process yields $N'$ estimates of the marginal likelihood, and $N' \times B'$ (correlated) samples drawn from the Markov chain.

To summarize, the overall AIS-MCMC sampling procedure corresponding to Equation~\ref{eqzs} is as follows:
\begin{itemize}
\item For each source type $a_s = a \in \{0, 1\}$ (e.g., ~star or galaxy)
\begin{itemize}
  \item Run $N'$ independent marginal likelihood estimators, each with $T$ annealing steps. This results in $N'$ independent estimates of $\log p(x | a_s=a, z_{-s})$ and $N'$ approximate posterior samples from $p(w_s | a_s = a, x, z_{-s})$.
  \item For each of the $N'$ approximate posterior samples, run an MCMC chain of length $B'$, using slice-sampling-within-Gibbs transitions.
\end{itemize}
\item Use the $\log p(x | a_s = 0, z_{-s})$ and $\log p(x | a_s=1, z_{-s})$ estimates to approximate $p(a_s = 1 | x, z_{-s})$.
\item Use the estimate of $p(a_s = 1 | x, z_{-s})$ to sample a source type $a_s^{(k)}$, approximating the distribution in Equation~\ref{eq:source-type-marginal}.
\item Randomly choose one of the $N' \times B'$ posterior samples corresponding to the realized $a_s^{(k)}$, which approximates the distribution $w_s^{(k)} \sim p(w_s | x, z_{-s}, a_s)$ from Equation~\ref{eq:source-param-conditional}; or collect all $N' \times B'$ samples to approximate posterior functionals.
\end{itemize}

The AIS-MCMC procedure described above requires us to choose a number of samples and iterations.  For the experiments we describe in Section \ref{sec:experiments}, we use $T = 200$ annealing steps and $N'=25$ independent samples of the marginal likelihood.  For each of the $N'$ samples, we run an additional slice-sampling MCMC chain for $B'=25$ iterations, producing a total of $N'\times B' = 625$ correlated posterior samples of $z_s$.


\section{Variational inference}
\label{sec:vi}

Variational inference (VI) chooses an approximation to the posterior
distribution $p(z | x)$ from a class of candidate distributions via numerical optimization.
The candidate approximating distributions $q_\theta(z)$, called ``variational distributions'', are parameterized by a real-valued vector $\theta$.
Through numerical optimization, VI minimizes (with respect to $\theta$) the KL divergence between $q_\theta(z)$ and $p(z|x)$.

For an introduction to VI, we recommend
\cite{blei2017variational} to statisticians,
\cite{mackay1995developments} to physicists, and
\cite{smidl2006variational} to readers with a background in signal processing.

\subsection{The variational distributions}
\label{vi_dist}

We restrict the variational distributions to a class that makes KL minimization tractable.
Our variational distributions all factorize:
\begin{align}
q_\theta(z) &=\prod_{s=1}^{S} q(a_{s}) q(u_{s}) q(e_{s}) q(r_{s}|a_{s}) q(c_{s}|a_{s}).\label{eq:q_factorization}
\end{align}
We have suppressed the subscript $\theta$ in the variational factors.
This is not quite mean-field variational inference \citep{blei2017variational},
where the variational distribution factorizes across all random variables,
because some factors are conditional on $a_s$ (i.e., whether a light source is a star or a galaxy).
The next equations show the constituents of $\theta$.
We use ``acute'' and ``hat'' accents to denote variational parameters.
For $s=1,\ldots,S$ and $i\in\{0,1\}$ we take
\begin{align}
q(a_{s}) & \sim\mathrm{Bernoulli}\left( \acute a_s \right),\label{eq0}\\
q\left(r_{s}|a_{s}=i\right) & \sim\mathrm{LogNormal}\left( \acute r_{si}, \hat r_{si}\right),\\
q\left(c_{s}|a_{s}=i\right) & \sim\mathrm{MvNormal}\left( \acute c_{si}, I \hat c_{si}\right),\\
q\left(u_{s}\right) & \sim\mathrm{PointMass}\left( \acute u_s \right),\\
q\left(e_{s}\right) & \sim\mathrm{PointMass}\left( \acute e_s \right).\label{eq1}
\end{align}
Here $\acute e_s \coloneqq (\acute e_s^{angle}, \acute e_s^{radius}, \acute e_s^{profile}, \acute e_s^{axis})$.

Approximating the posterior for $u_s$ and $e_s$ with a point mass is a strong assumption.
It is analogous to performing maximum a posteriori (MAP) inference for these parameters.
We do so only because of computational considerations: it lets us write the objective function as an analytic expression.
Analytic expressions can be optimized efficiently by deterministic numerical optimization routines, which
in turn can converge much faster than stochastic optimization \citep{bubeck2015convex}.
Ongoing research aims to expand the class of models and variational distributions that can be optimized with deterministic VI,
though limitations persist~\citep{fraysse2014measure,zheng2015efficient,giordano2015linear}.

\subsection{The variational lower bound}

Because $p(x)$ is constant with respect to $\theta$,
minimizing $D_{\mathrm{KL}}(q_\theta(z), p(z|x))$ is equivalent to maximizing
\begin{align}
\mathcal L(\theta) \coloneqq \mathbb{E}_{q_\theta} \left[ \log p(x | z) \right]
                   - D_{KL}(q_\theta(z), p(z)).\label{elbo}
\end{align}
Maximization of $\mathcal L(\theta)$ is the standard approach;
see \cite{blei2017variational} for discussion.

The first term of $\mathcal L(\theta)$ is the expected log likelihood of the data.
It is
\begin{align}
\mathbb{E}_{q} \left[\log p(x|z)\right]
&=\sum_{n=1}^{N} \sum_{m=1}^{M} \left\{
  -\mathbb{E}_{q}\left[\lambda_{nm}\right]
  + x_{nm}\mathbb{E}_{q}\left[\log \lambda_{nm}\right]
  -\log\left(x_{nm}!\right)
\right\}.\label{eq:log_p_last}
\end{align}

\subsubsection{Expectation of the rate parameter}
The first expectation is
\begin{align}
\mathbb{E}_{q}\left[\lambda_{nm}\right]
&= \sigma_{nm} + \sum_{s=1}^S
     \mathbb E_q \left[ \ell_{s\beta_n} f_{nms} \right].
\end{align}
We can factorize the right-hand expectation based on the factorization of the variational distribution, upon conditioning on $a_s$:
\begin{align}
\begin{split}
\mathbb E_q \left[ \ell_{s\beta_n} f_{nms} \right]
= (1 - \acute a_s) &\mathbb E_q \left[ \ell_{s\beta_n} | a_s = 0 \right]\mathbb E_q \left[ f_{nms} | a_s = 0 \right]\\
+ \acute a_s &\mathbb E_q \left[ \ell_{s\beta_n} | a_s = 1 \right]\mathbb E_q \left[ f_{nms} | a_s = 1 \right].
\end{split}
\label{conditional}
\end{align}
The integral $\mathbb E_q \left[ \ell_{s\beta} | a_s \right]$ is tractable because flux~$r_s$
and each entry of~$c_s$ (the colors) are independent in the variational distribution given $a_s$.
The integral~$\mathbb E_q \left[ f_{nms} | a_s \right]$ is tractable because $u_s$ is a point mass in the variational distribution.

\subsubsection{Expectation of the log rate parameter}

We approximate the expected logarithm of $\lambda_{nm}$ using the delta method for moments \citep{bickel2015mathematical}.
We replace the integrand with a second-order Taylor expansion around its mean:
\begin{align}
\begin{split}
\log(\lambda_{nm})
\approx \log\mathbb{E}_{q}[\lambda_{nm}]
  &+\frac{1}{\mathbb{E}_{q}[\lambda_{nm}]}\left(\lambda_{nm} - \mathbb{E}_{q}[\lambda_{nm}]\right)\\
  &-\frac{1}{2\mathbb{E}_{q}[\lambda_{nm}]^{2}}\left(\lambda_{nm}-\mathbb{E}_{q}[\lambda_{nm}]\right)^{2}\,.
\end{split}
\end{align}
Then, taking expectations,
\begin{align}
\mathbb{E}_{q}[ \log(\lambda_{nm}) ]
&\approx \log\mathbb{E}_{q}[\lambda_{nm}]
  -\frac{\mathbb{V}_{q}[\lambda_{nm}]}{2\mathbb{E}_{q}[\lambda_{nm}]^{2}},
\end{align}
where $\mathbb V_q$ denotes variance with respect to the variational distribution $q$.
That term may be further expanded:
\begin{align}
\mathbb{V}_{q}[\lambda_{nm}] & =\sum_{s=1}^{S}\mathbb{V}_{q}\left[\ell_{s\beta_n}f_{nms}\right]\\
&=\sum_{s=1}^{S} \mathbb{E}_{q}\left[\ell_{s\beta_n}^{2}f_{nms}^{2}\right]
    - \left(\mathbb{E}_{q}\left[\ell_{s\beta_n}f_{nms}\right]\right)^{2}.
    \label{vareq}
\end{align}
The second expectation on the right-hand side is given in Equation~\ref{conditional}. The first is
\begin{align}
\begin{split}
\mathbb{E}_{q}\left[\ell_{s\beta_n}^{2}f_{nms}^{2}\right]
&= (1 - \acute a_s) \mathbb E_q \left[ \ell_{s\beta_n}^2 | a_s = 0 \right]\mathbb E_q \left[ f_{nms}^2 | a_s = 0 \right]\\
&+ \acute a_s \mathbb E_q \left[ \ell_{s\beta_n}^2 | a_s = 1 \right]\mathbb E_q \left[ f_{nms}^2 | a_s = 1 \right].
\end{split}
\end{align}

\subsubsection{KL divergence}

Because of the factorization of the variational distribution, the KL term in Equation~\ref{elbo} separates across sources:
\begin{align}
D_{\mathrm{KL}}(q(z), p(z)) &= \sum_{s=1}^S D_{\mathrm{KL}}(q(z_s), p(z_s)).\label{eq:kl}
\end{align}
It separates further within each source:
\begin{align}
\begin{split}
D_{\mathrm{KL}}(q(z_s), p(z_s)) &=
D_{\mathrm{KL}}(q(a_s), p(a_s))\\
&+ D_{\mathrm{KL}}(q(u_s), p(u_s))
+ D_{\mathrm{KL}}(q(e_s), p(e_s))\\
&+ \sum_{i=0}^1 q(a_s = i) \Big[
    D_{\mathrm{KL}}(q(r_s | a_s = i), p(r_s | a_s = i))\\
    &\qquad\qquad\qquad + D_{\mathrm{KL}}(q(c_s | a_s = i), p(c_s | a_s = i))
\Big].
\end{split}
\label{kleq}
\end{align}
Except for the last, these KL divergences are between common exponential family distributions. We give formulas for them in~\ref{kl}.

The last KL divergence is more complicated because the prior on $c_s$ is a Gaussian mixture model.
We take the eighth approach from \citet{hershey2007approximating} to identify an upper bound on this KL divergence:
\begin{multline}
D_{\mathrm{KL}}(q(c_s | a_s = i), p(c_s | a_s = i)) \\ \le
D_{\mathrm{KL}}(\xi_i, \mathcal C_{i}^{weights}) + \sum_{j=1}^J \xi_{ij} D_{\mathrm{KL}}(q(c_s | a_s = i), \mathcal C_{ij}).
\end{multline}
Here $\mathcal C_{i}^{weights}$ is the categorical distribution over the color prior's mixture components, $\mathcal C_{ij}$ is the color prior's $j$th mixture component, and $\xi_{i} \in [0,1]^J$ is a vector of free parameters.
To make the bound as tight as possible, we optimize the $\xi_i$ along with the variational lower bound. The optimal $\xi_i$
can also be expressed analytically in terms of $\mathcal C_i$:
\begin{align}
\xi_{ij}^\star \propto \mathcal C_{ij}^{weights} \exp\{-D_{\mathrm{KL}}(q(c_s | a_s = i), \mathcal C_{ij})\}.
\end{align}

\subsection{Numerical optimization}
\label{numerical}

Traditionally, variational lower bounds are maximized through coordinate ascent: each update sets a variational parameter to its optimal value with the others held fixed~\citep{bishop2006pattern,murphy2012machine}. This approach is simple to implement because gradients and Hessians do not need to be explicitly computed. Each update increases the variational lower bound. The algorithm converges to a local optimum even for nonconvex objective functions.
However, coordinate ascent can take many iterations to converge when the Hessian of the objective function is not diagonal.
Additionally, for many models, including ours, optimal coordinate ascent updates cannot be expressed analytically.

Instead, we propose an optimization procedure based on \textit{block} coordinate ascent. Each light source corresponds to a block of 44 parameters: the 37 variational parameters in Equations~\ref{eq0}--\ref{eq1} and the $7$-dimensional parameter~$\xi$. We optimize each block using a subsolver, explained in the next paragraph. Because most pairs of light sources do not overlap, the Hessian has low fill off the block diagonal.
Block coordinate ascent converges quickly in this setting: for light sources that do not overlap with any other light source, just one update step, based on one call to a subsolver, is required to reach a local maximum. For groups of light sources that overlap with each other, a few passes over each light source suffice in practice. Light sources may be optimized in a round-robin order or at random.

As a subsolver to optimize one block of parameters with all others fixed, we use Newton's method with a trust-region constraint that restricts each step to a Euclidean ball centered at the previous iterate~\citep{nocedal2006numerical}.
The trust-region constraint ensures that we find a local maximum even though the variational objective is nonconvex.
The method consistently converges in tens of iterations, whereas first-order methods take thousands. BFGS~\citep{nocedal2006numerical} also on occasion required thousands of iterations per call. Newton iterations are more expensive computationally than the iterations of first-order methods  because the former require computing a dense Hessian along with each gradient.
For our objective function, computing both a Hessian and a gradient takes $3\times$ longer than computing a gradient alone. In the end, we gain at least an order of magnitude speedup by using Newton's method rather than a gradient-only method because the former requires many fewer iterations.

\subsection{Distributed optimization}
\label{distributed}

Modern compute clusters and supercomputers contain many individual \textit{compute nodes} that execute instructions in parallel. Additionally, each compute node runs many \textit{threads} in parallel---at least one per CPU core.
Communication among compute nodes is orders of magnitude slower than communication among threads on the same node.

Block coordinate ascent (the outer loop of our optimization procedure) is a serial algorithm: if multiple blocks of parameters are updated simultaneously based on the current iterate, the objective value may decrease, and the algorithm may diverge. By taking of advantage of the structure of our problem, however, we parallelize block coordinate ascent across both compute nodes and CPU cores.
Equation~\ref{eq:log_p_last} is a sum over pixels and Equation~\ref{eq:kl} is a sum over light sources.
Therefore, our objective function may be expressed as a sum whose terms each depend on the parameters for at most one light source from any particular collection of non-overlapping light sources.
Thus, for any collection of non-overlapping light sources, maximizing over each light source's parameters serially is equivalent to maximizing over all these light sources' parameters in parallel.

Each compute node is tasked with optimizing all the light sources in a region of the sky.
Because these light sources are physically near each other, they appear in many of the same images; we only need to load these images once for to infer parameters for all these light sources.
Each node implements a locking mechanism that prevents its threads from optimizing overlapping light sources simultaneously. Because within-node communication is fast, there is almost no overhead from this type of locking mechanism.

Communication between nodes is relatively slow. We avoid using an inter-node locking mechanism by assigning each node to optimize different regions of the sky. Because the boundaries of these regions are small relative to the interior,
we find an iterate near a stationary point with this approach. A second pass with shifted boundaries ensures that even light sources near a boundary during the first pass are fully optimized.

\section{Experimental results}
\label{sec:experiments}
Our experiments aim to assess
1)~how MCMC and VI compare, statistically and computationally; and
2)~how well our procedures quantify uncertainty.

We base our experiments both on synthetic images drawn from our model (Section~\ref{sec:synth}) and images from the Sloan Digital Sky Survey (Section~\ref{sec:real}).
For both datasets, we run both
the MCMC procedure from Section~\ref{sec:mcmc} (henceforth, MCMC)
and the variational inference procedure from Section~\ref{sec:vi} (henceforth, VI),
and compare their posterior approximations.%
\footnote{Open-source software implementing our inference procedures is available from \url{https://github.com/jeff-regier/Celeste.jl}. Jupyter notebooks demonstrating how to replicate all reported results are stored in the \texttt{experiments} directory.}

We assess the accuracy of point estimates (e.g., posterior means/modes) and uncertainties (e.g., posterior variances), as well as star/galaxy classification accuracy.
Our accuracy measures are averaged over a population of light sources.
While no single metric of quality suffices for all downstream uses of catalogs,
good performance on the metrics we report is necessary (though not sufficient) for good performance on most downstream tasks.
These error metrics, which are an unweighted average across light sources, are likely more representative of performance at spectrograph targeting than for demographic inference. (Demographic inference and spectrograph targeting are the two downstream applications we introduced in Section~\ref{sec:intro}.)

\subsection{Synthetic images}
\label{sec:synth}

\begin{figure}[!hb]
\begin{subfigure}{.49\textwidth}
\includegraphics[width=0.95\textwidth]{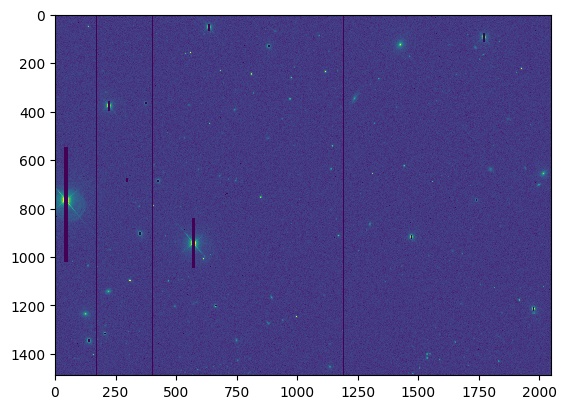}
\end{subfigure}
\begin{subfigure}{.49\textwidth}
\includegraphics[width=0.95\textwidth]{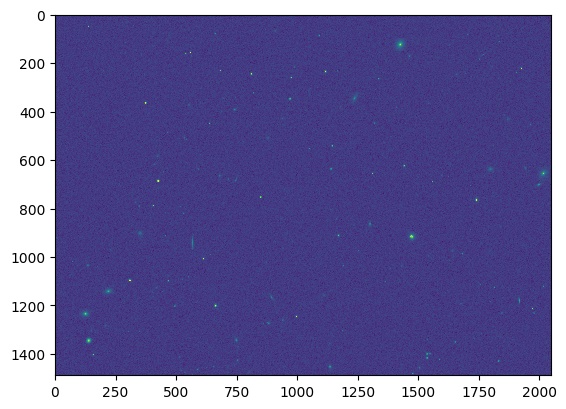}
\end{subfigure}
\caption{\textit{Left:} An image from SDSS containing approximately 1000 detectable light sources.
Pixels in error are ``masked'' (black strips).
\textit{Right:} A synthetic image for the same region, generated from our model by conditioning on an SDSS catalog for that region. (Several of the light sources with extremely high flux are excluded---the CCDs cannot record such high flux.)}
\label{real-synthetic}
\end{figure}

\begin{figure}
\begin{floatrow}
\ffigbox{%
  \includegraphics[width=\linewidth]{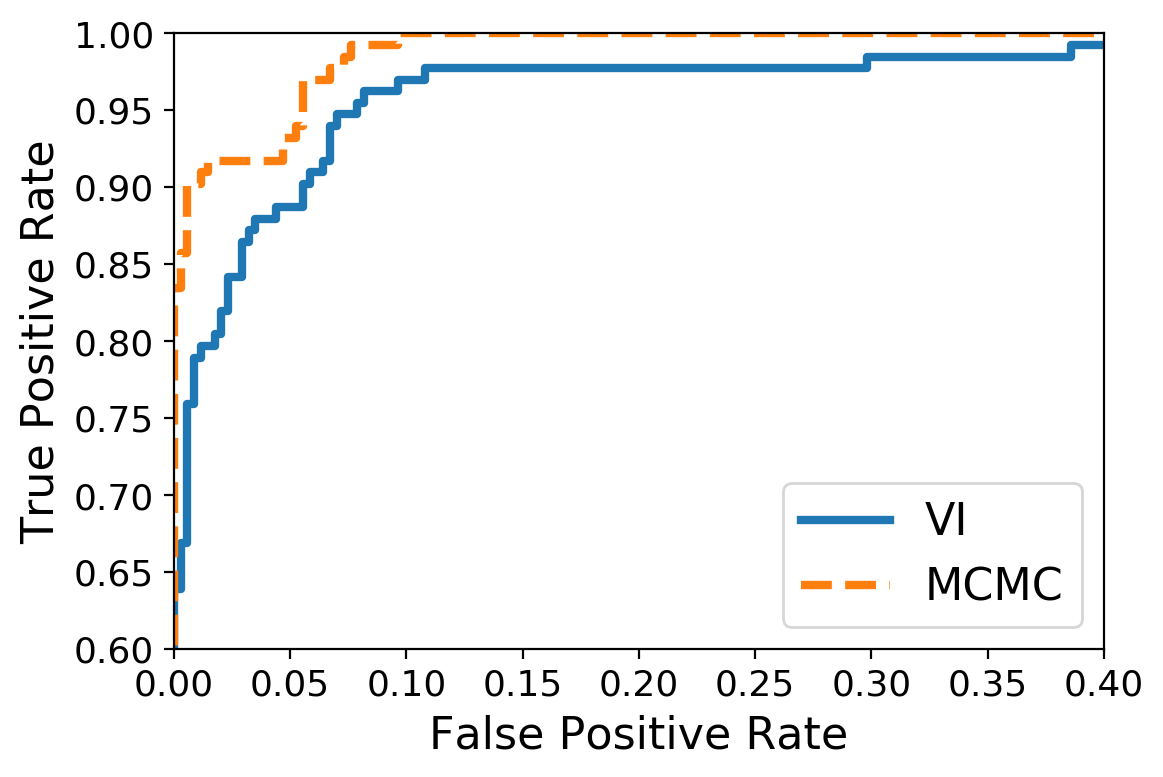}
}{%
  \caption{ROC curve for star/galaxy classification on synthetic data.}
  \label{roc-synth}
}
\capbtabbox{%
\scalebox{.85}{
\begin{tabular}{l|ll|rl}
\toprule
{} &   MCMC &     VI &                \multicolumn{2}{|c}{VI-MCMC} \\
\midrule
direction   &  0.111 &  0.121 &   \textbf{0.010} & \hspace{-1em}($\pm$ 0.003) \\
flux &  0.093 &  0.118 &   \textbf{0.025} & \hspace{-1em}($\pm$ 0.006) \\
color u-g  &  0.327 &  0.333 &   0.006          & \hspace{-1em}($\pm$ 0.008) \\
color g-r  &  0.128 &  0.126 &  -0.002          & \hspace{-1em}($\pm$ 0.004) \\
color r-i  &  0.112 &  0.110 &  -0.002          & \hspace{-1em}($\pm$ 0.005) \\
color i-z  &  0.154 &  0.144 &  -0.010          & \hspace{-1em}($\pm$ 0.005) \\
galaxy profile &  0.158 &  0.229 & \textbf{0.072} & \hspace{-1em}($\pm$ 0.011) \\
galaxy axis    &  0.074 &  0.106 & \textbf{0.032} & \hspace{-1em}($\pm$ 0.006) \\
galaxy radius  &  0.450 &  0.688 & \textbf{0.237} & \hspace{-1em}($\pm$ 0.043) \\
galaxy angle   &  9.642 &  8.943 & -0.699         & \hspace{-1em}($\pm$ 0.437) \\
\bottomrule
\end{tabular}

}
\vspace{0.5em}
}{%
\caption{%
Left columns: Mean absolute error on synthetic data.
Right column:~Pairwise error differences (and standard error).
Statistically significant differences appear in bold font.}
\label{tab:err-synth}
}
\end{floatrow}
\end{figure}

\begin{figure}
\centering
\begin{subfigure}{.49\textwidth}
  \centering
  \includegraphics[width=\linewidth]{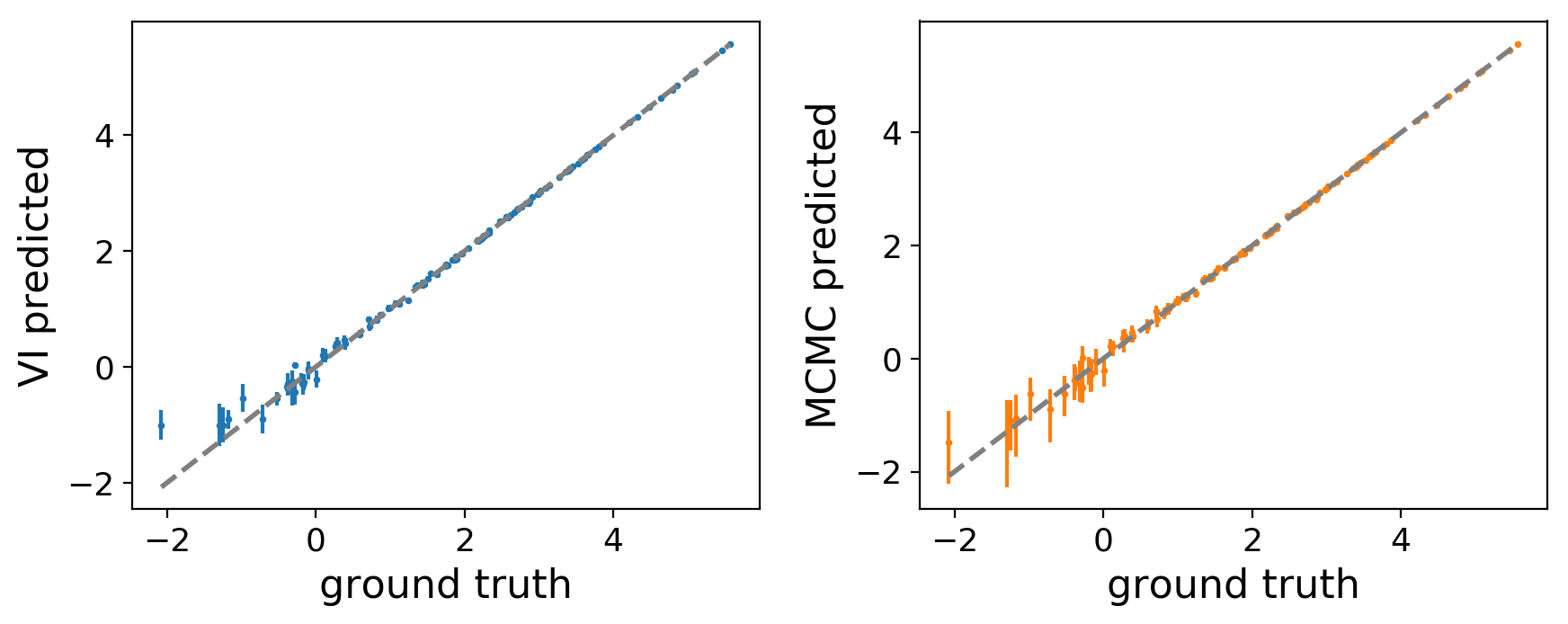}
  \caption{log flux (stars)}
\end{subfigure}
\begin{subfigure}{.49\textwidth}
  \centering
  \includegraphics[width=\linewidth]{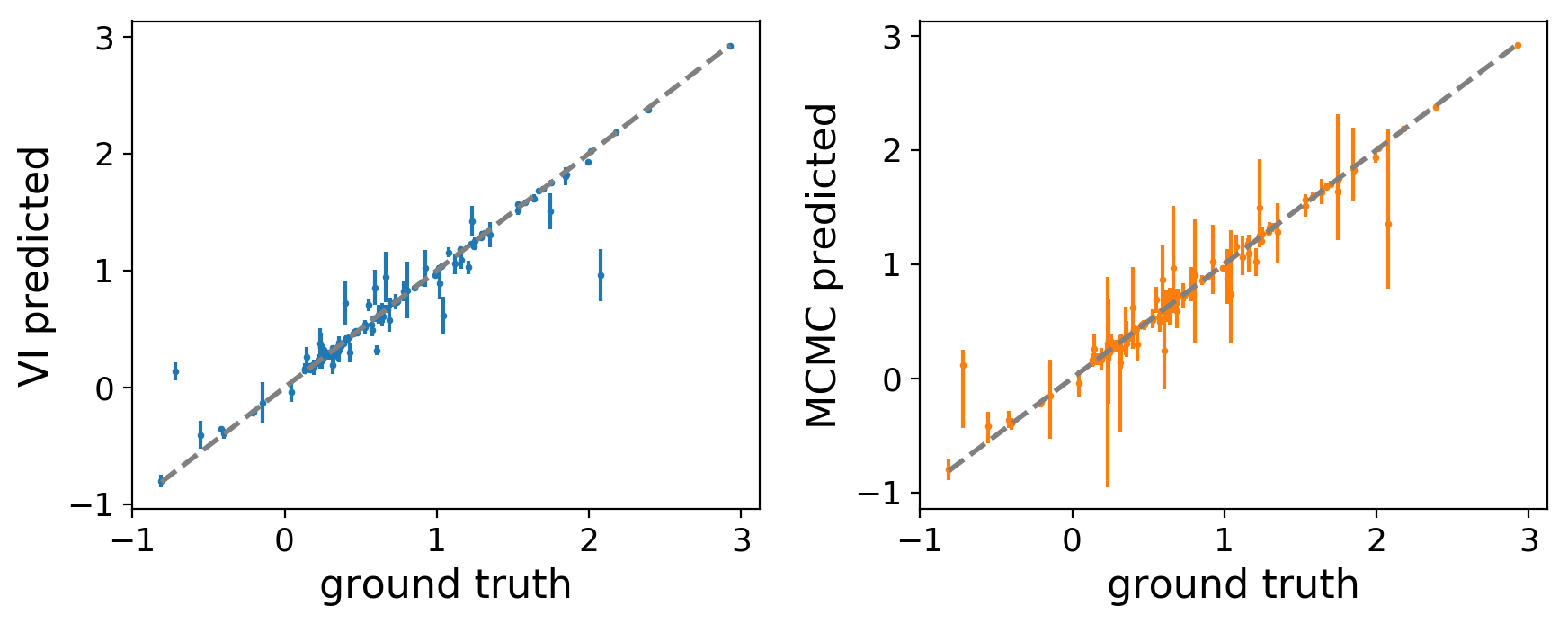}
  \caption{color r-i (stars)}
\end{subfigure}
\\
\begin{subfigure}{.49\textwidth}
  \centering
  \includegraphics[width=\linewidth]{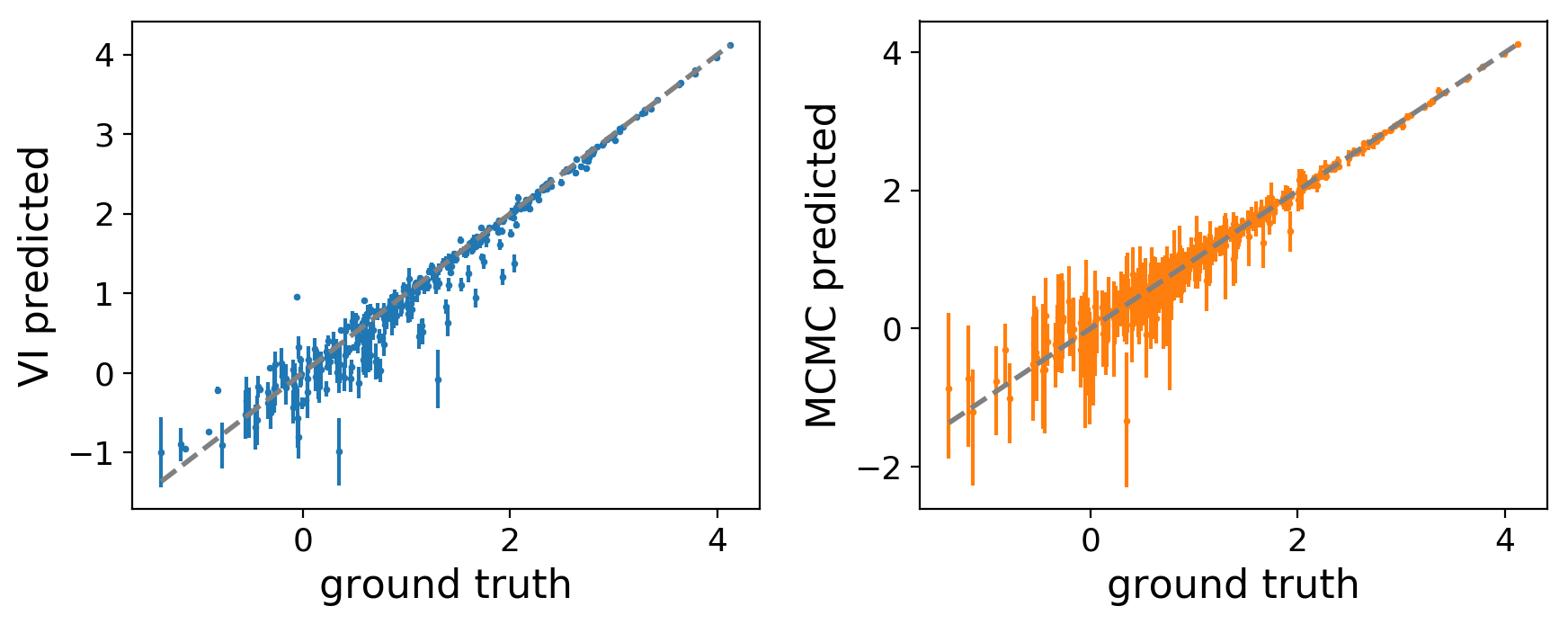}
  \caption{log flux (galaxies)}
\end{subfigure}
\begin{subfigure}{.49\textwidth}
  \centering
  \includegraphics[width=\linewidth]{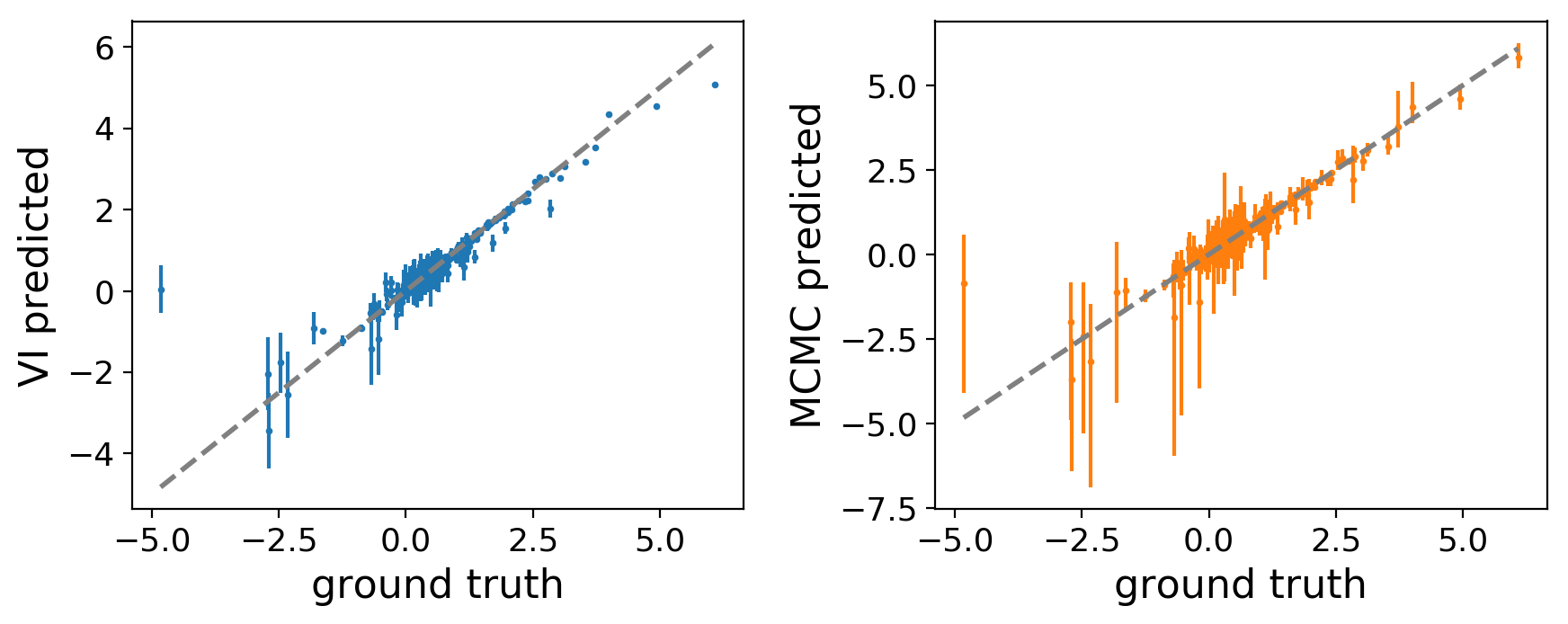}
  \caption{color r-i (galaxies)}
\end{subfigure}
\caption{VI and MCMC performance on synthetic data.  Each pair depicts VI (left, blue) and MCMC (right, orange) with the ground truth along the $x$-axis and the posterior distribution (showing equal-tailed 95.4\% credible intervals) along the $y$-axis.}
\label{qq-synth}
\end{figure}

\begin{table}[b]
\scalebox{.75}{
\begin{tabular}{lrrrr}
{} &  &  \textbf{VI} &  & \\
\toprule
{} & within 1/2 sd &  1 sd &  2 sd &  3 sd \\
\midrule
log flux &         0.18 & 0.31 & 0.55 & 0.68 \\
color u-g   &         0.29 & 0.52 & 0.79 & 0.89 \\
color g-r   &         0.26 & 0.46 & 0.71 & 0.80 \\
color r-i   &         0.22 & 0.43 & 0.72 & 0.84 \\
color i-z   &         0.32 & 0.58 & 0.82 & 0.93 \\
\bottomrule
\end{tabular}

}~
\scalebox{.75}{
\begin{tabular}{lrrrr}
{} &  &  \textbf{MCMC} &  & \\
\toprule
{} & within 1/2 sd & 1 sd & 2 sd & 3 sd \\
\midrule
log\_flux\_r &          0.35 & 0.63 & 0.91 & 0.98 \\
color ug   &          0.40 & 0.71 & 0.94 & 0.99 \\
color gr   &          0.38 & 0.65 & 0.93 & 0.99 \\
color ri   &          0.38 & 0.65 & 0.93 & 0.99 \\
color iz   &          0.37 & 0.67 & 0.95 & 0.99 \\
\bottomrule
\end{tabular}

}
\vspace{.5em}
\caption{
Proportion of light sources having posterior means found by VI (left) and MCMC (right) near the ground truth for synthetic images.
The VI credible intervals correspond to the estimated posterior standard deviation.
For MCMC, we match these with equal-tailed credible intervals derived from samples, where one-half standard deviation (sd) covers 38.2\% of probability mass, 1 sd covers 68.3\%, 2 sds covers 95.4\% and 3 sds covers 99.7\%.}
\label{tab:calibration-synth}
\end{table}

\begin{figure}[b]
\begin{subfigure}{.32\textwidth}
  \centering
  \includegraphics[width=\linewidth]{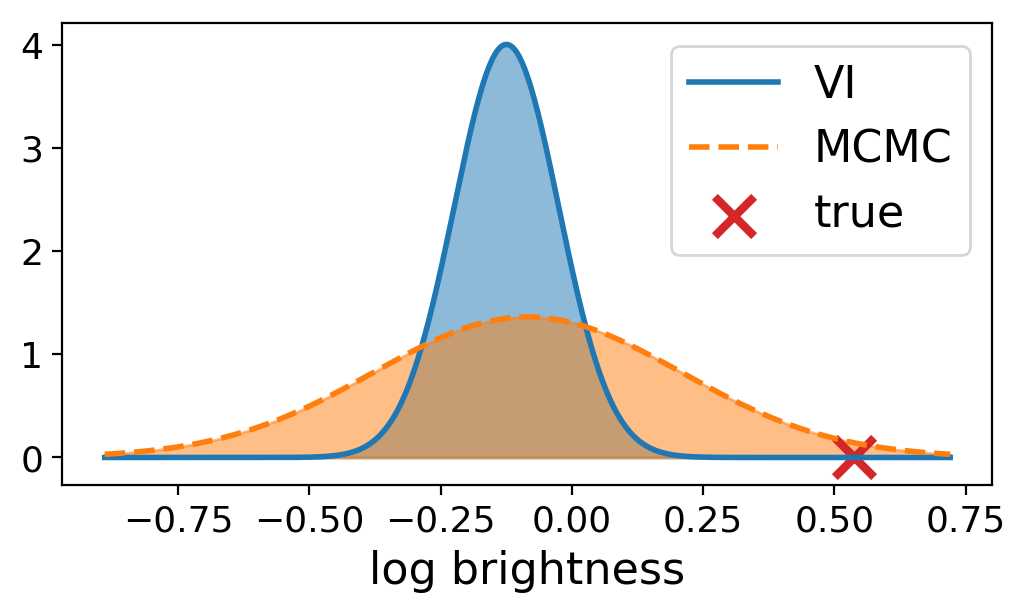}
  \caption{}
\end{subfigure}
\begin{subfigure}{.32\textwidth}
  \centering
  \includegraphics[width=\linewidth]{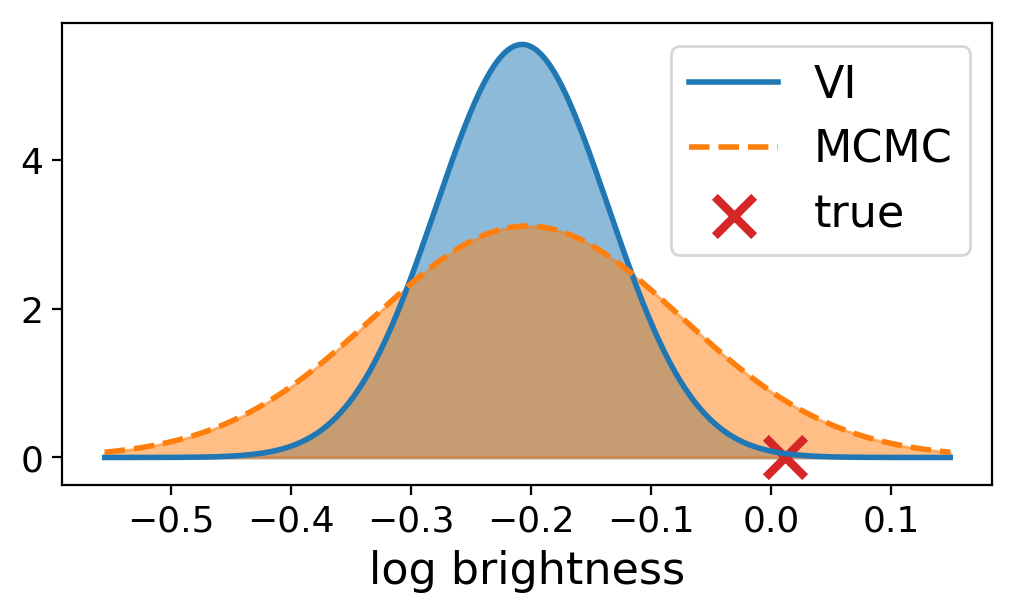}
  \caption{}
\end{subfigure}
\begin{subfigure}{.32\textwidth}
  \centering
  \includegraphics[width=\linewidth]{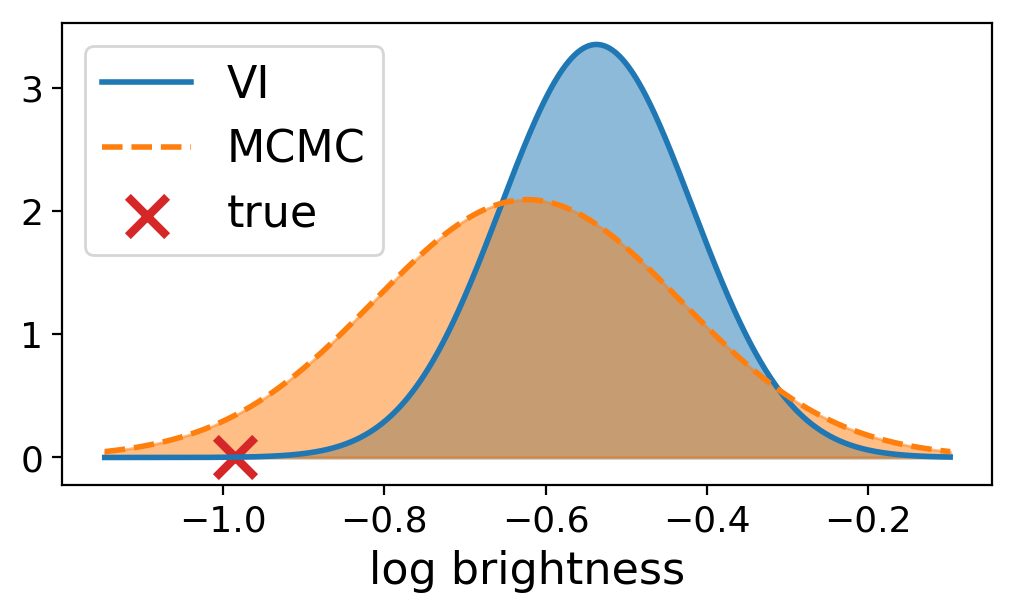}
  \caption{}
\end{subfigure}
  \caption{Comparison of posterior uncertainty for the flux of three synthetic light sources where the posterior mean is a poor prediction of the true parameter value. VI underestimates posterior uncertainty. MCMC assigns much greater posterior density to the true values.}
\label{uq-synth}
\end{figure}

Synthetic images let us compare inference methods without model misspecification.
On synthetic images, ``ground truth'' for the latent random variables is known.
Synthetic images also let us validate our model by visually checking their similarity to real images.
To generate realistic synthetic images, we take the non-inferred parameter values from real SDSS images, including the point-spread function $\psi_n$, the sky background $\sigma_n$, and structural constants like the dimensions of the images.
To illustrate that synthetic data resemble real images, Figure~\ref{real-synthetic} depicts a synthetic image generated using parameters from an existing catalog.
In our experiments, the light sources in synthetic images are instead drawn from the prior.
Our synthetic study set comprises five overlapping $2048 \times 1489$-pixel images.
Each image is for a different filter band.
The images contain approximately 500 detectable light sources.

Empirically, MCMC performs better for star/galaxy classification than VI for all thresholds of a receiver operating characteristic (ROC) curve (Figure~\ref{roc-synth}).
Both methods have a high area under the curve (AUC).
For MCMC, the AUC is 0.994. For VI, the AUC is 0.981.

Both methods estimate means well for all continuous latent random variables (Table~\ref{tab:err-synth}).
MCMC outperforms VI significantly for some point estimates.
``Direction'' is error, in arcseconds (0.396 pixels), for the directions of the light sources'
centers.
``Flux'' measures the reference band (r-band) flux.
``Colors'' are ratios of fluxes in consecutive bands.
``Galaxy profile'' is a proportion indicating whether a galaxy is de Vaucouleurs
or exponential.
``Galaxy axis'' is the ratio between the lengths of a galaxy's minor and
major axes.
``Galaxy radius'' is the half-light radius of a galaxy in arcseconds.
``Galaxy angle'' is the orientation of a galaxy in degrees.

For color and flux, MCMC often has larger posterior uncertainty.
MCMC assigns substantial probability density to the truth more often than VI (Figure~\ref{qq-synth}).
For light sources where posterior means are particularly poor predictors of the truth, VI severely underestimates the uncertainty, whereas MCMC assigns much greater posterior density to the true values (Figure~\ref{uq-synth}).
For color and log flux---both normally distributed quantities in this synthetic data---errors from MCMC are more nearly normally distributed than those of VI.
Table~\ref{tab:calibration-synth} reports the fraction of sources covered by equal-tailed posterior credible intervals of increasing width. The MCMC uncertainty estimates are more accurately calibrated.
The typical range of effectively independent samples generated MCMC is between 100 and 150 per source.
For a single source, 140 samples is sufficient to approximate a 60\% credible interval with high probability \citep{booth1998monte}.
However, we note that we are averaging over 500 sources, each with independent samples, allowing us to resolve population posterior coverage with higher fidelity.

These empirical results are anticipated by theory: VI underestimates the posterior uncertainty because independence assumptions in the variational distribution do not hold in the posterior~\citep{bishop2006pattern}. Additionally, differences between the candidate variational distributions' marginals and the posteriors' marginals are a source of bias. For the marginals we approximate with point masses (those of $u_s$ and $e_s$), that may be a particularly important source of bias.

\subsection{Real images from SDSS}
\label{sec:real}

Absolute truth is not currently knowable for astronomical catalogs. Fortunately, one area of the sky, called ``Stripe 82,'' has been imaged many times in SDSS.
This region provides a convenient validation strategy: combine exposures from all Stripe-82 runs to produce a high signal-to-noise image, then use parameters estimated from the combined exposure as a surrogate ground truth.

Photo~\citep{lupton2005sdss} is the primary software pipeline for cataloging SDSS.
We use Photo's estimated parameters from the combined Stripe 82 imagery as ground truth.
We then run Photo and our method on just one of the 80 image sets, comparing the results from each to the ground truth.

To reduce the runtime of our algorithms, we test them on only a subset of Stripe 82.
Our Stripe 82 study set comprises five overlapping $2048 \times 1489$-pixel images for a typical region of sky.
Each of these images is captured through a different filter.
The images contain approximately 500 detectable light sources.

\begin{figure}
  \centering
  \begin{subfigure}{.45\textwidth}
    \centering
    \includegraphics[width=\linewidth]{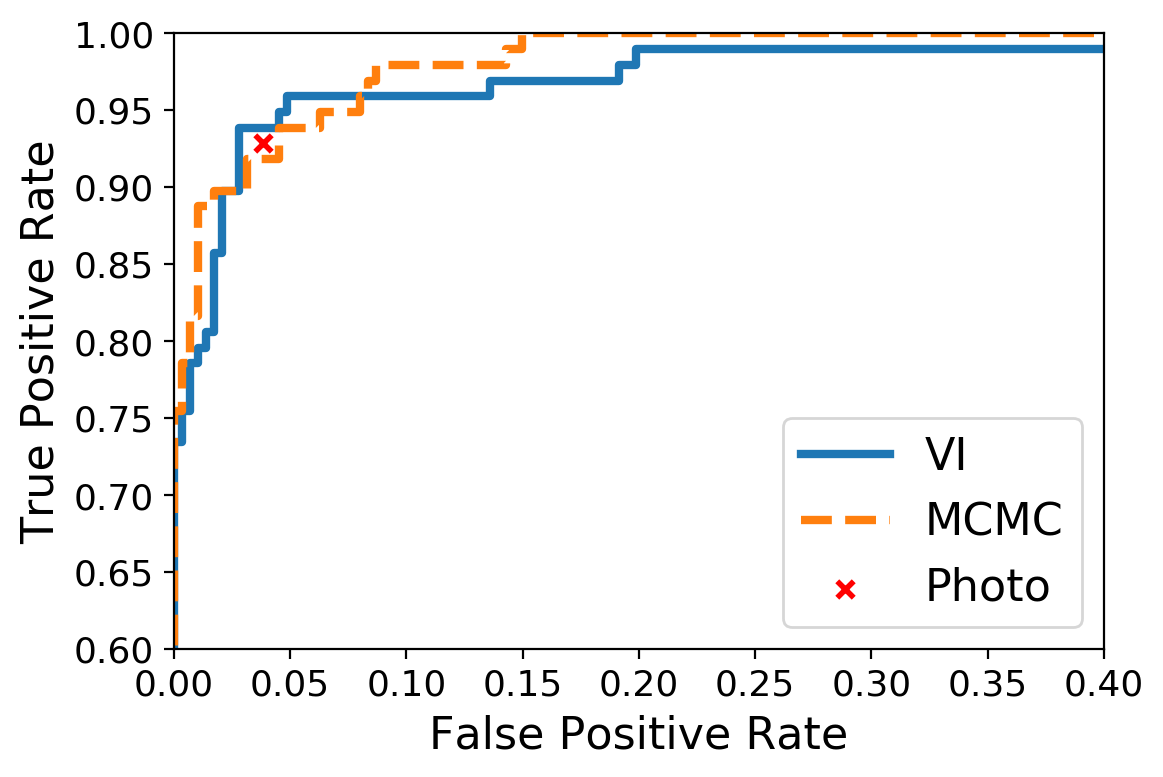}
  \end{subfigure}
  \caption{The receiver operating characteristic (ROC) curve for star/galaxy classification on Stripe 82 data. The area under the curve (AUC) for MCMC is 0.991 and for VI is 0.985.
  }
\label{roc-real}
\end{figure}

For star/galaxy classification in SDSS data, MCMC outperforms VI at some thresholds and performs slightly worse than VI at others (Figure~\ref{roc-real}). In addition to point estimates, our inference procedures approximate posterior uncertainty for source type (star or galaxy), flux, and colors. This is a novel feature of a Bayesian approach, offering astronomers a principled measure of the quality of inference for each light source; Photo gives only conditional uncertainty estimates.

The MCMC procedure is certain ($> 99\%$ certainty) about the classification (star vs. galaxy) for 321 out of 385 light sources.
Of these classifications, 319 (99.4\%) are correct. Of the remaining classifications (>1\% uncertainty), 50 (78.1\%) are correct.
The VI procedure is certain ($> 99\%$ certainty) about the classification for 322 out of 385 light sources.
Of these classifications, 318 (98.8\%) are correct. Of the remaining classifications (>1\% uncertainty), 53 (84.1\%) are correct.

Table~\ref{tab:err-s82} quantifies point-estimate error from MCMC and VI for the real-valued latent random variables, as well as providing a paired error comparison between each method.
Point-estimate errors for MCMC and VI differed significantly only for galaxy profile and galaxy axis ratio.
For galaxy axis, MCMC outperformed VI, repeating our experience with synthetic data.
For galaxy profile, however, VI outperformed MCMC---the opposite of how the methods compared on synthetic data.
Sampler diagnostics, though not conclusive, suggest that insufficient mixing was not to blame.
Model misfit, though an obvious explanation for any result not shared by synthetic data, seems inadequate because MCMC recovered the other galaxy shape parameters at least as well as VI.

Our leading explanation is that ``ground truth'' is unreliable for galaxy profile, and that VI more accurately recreates the ground-truth mistakes.
Recall ground truth is determined by additively combining many overlapping images.
These images were taken through a variety of atmospheric conditions.
Errors in the point-spread function (PSF) are likely compounded by the addition of more data.
Galaxy profile may be particularly susceptible to errors in the PSF because it has the capacity to model image blur that should have been attributed to the PSF.

\begin{figure}
\centering
\begin{subfigure}{.49\textwidth}
  \centering
  \includegraphics[width=\linewidth]{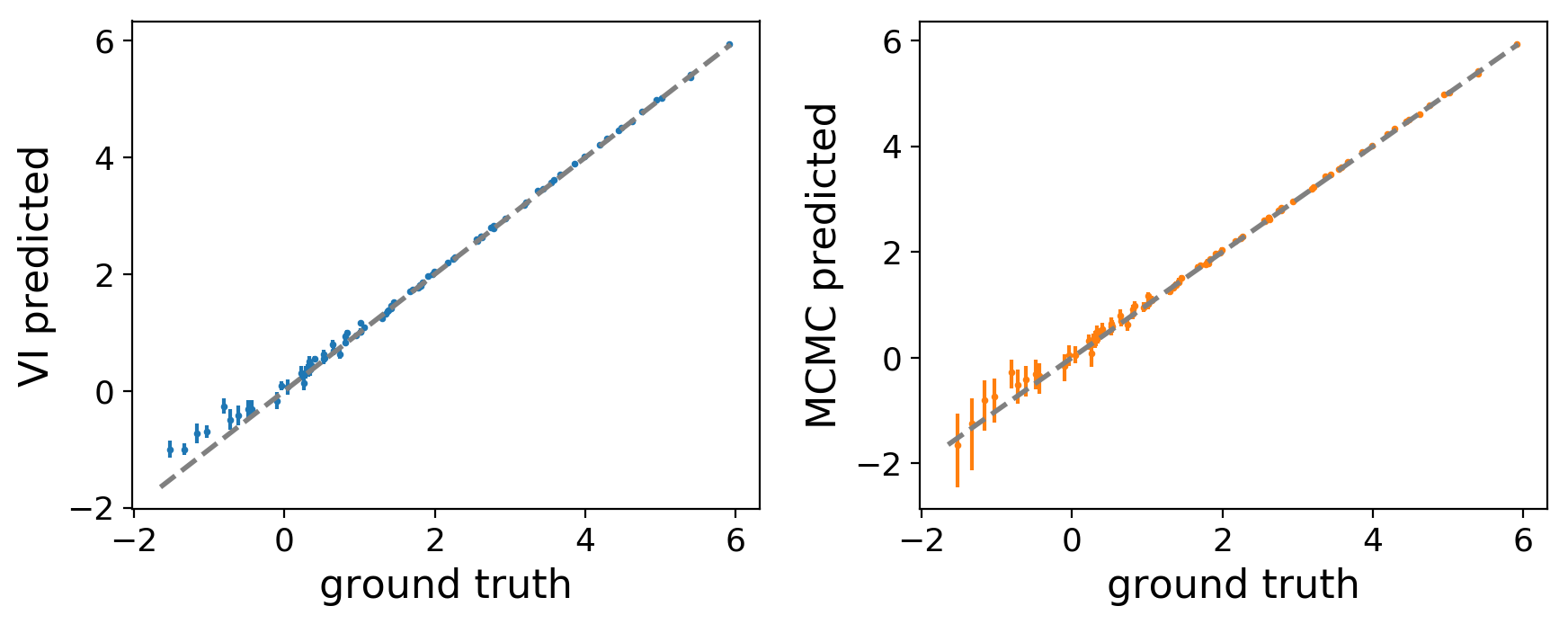}
  \caption{log flux (stars)}
\end{subfigure}
\begin{subfigure}{.49\textwidth}
  \centering
  \includegraphics[width=\linewidth]{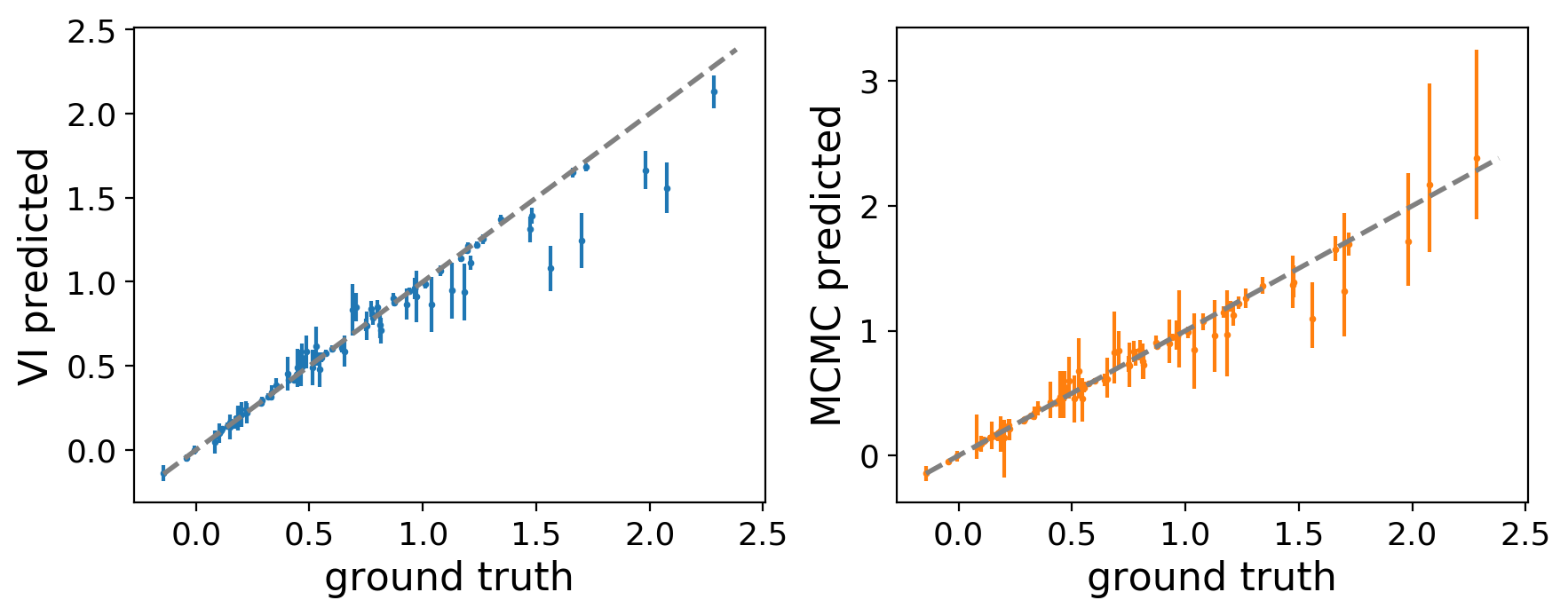}
  \caption{color r-i (stars)}
\end{subfigure}
\\
\begin{subfigure}{.49\textwidth}
  \centering
  \includegraphics[width=\linewidth]{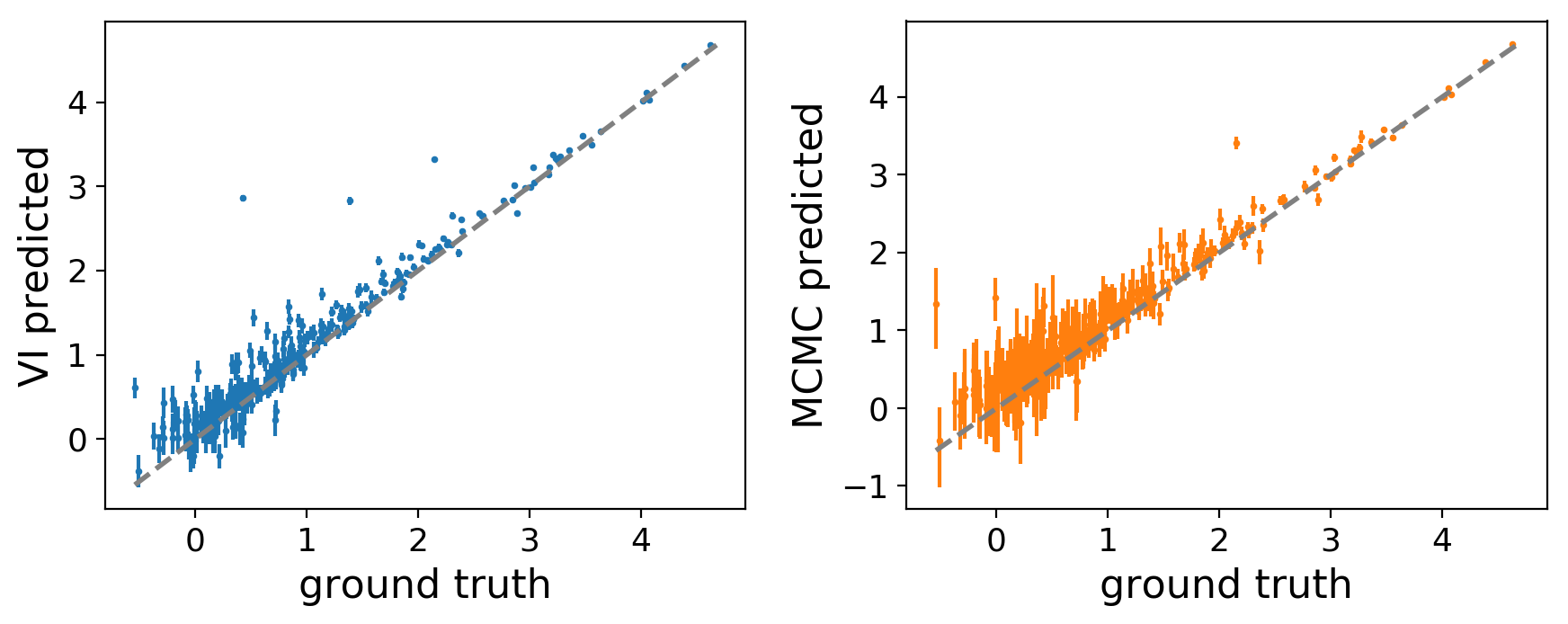}
  \caption{log flux (galaxies)}
\end{subfigure}
\begin{subfigure}{.49\textwidth}
  \centering
  \includegraphics[width=\linewidth]{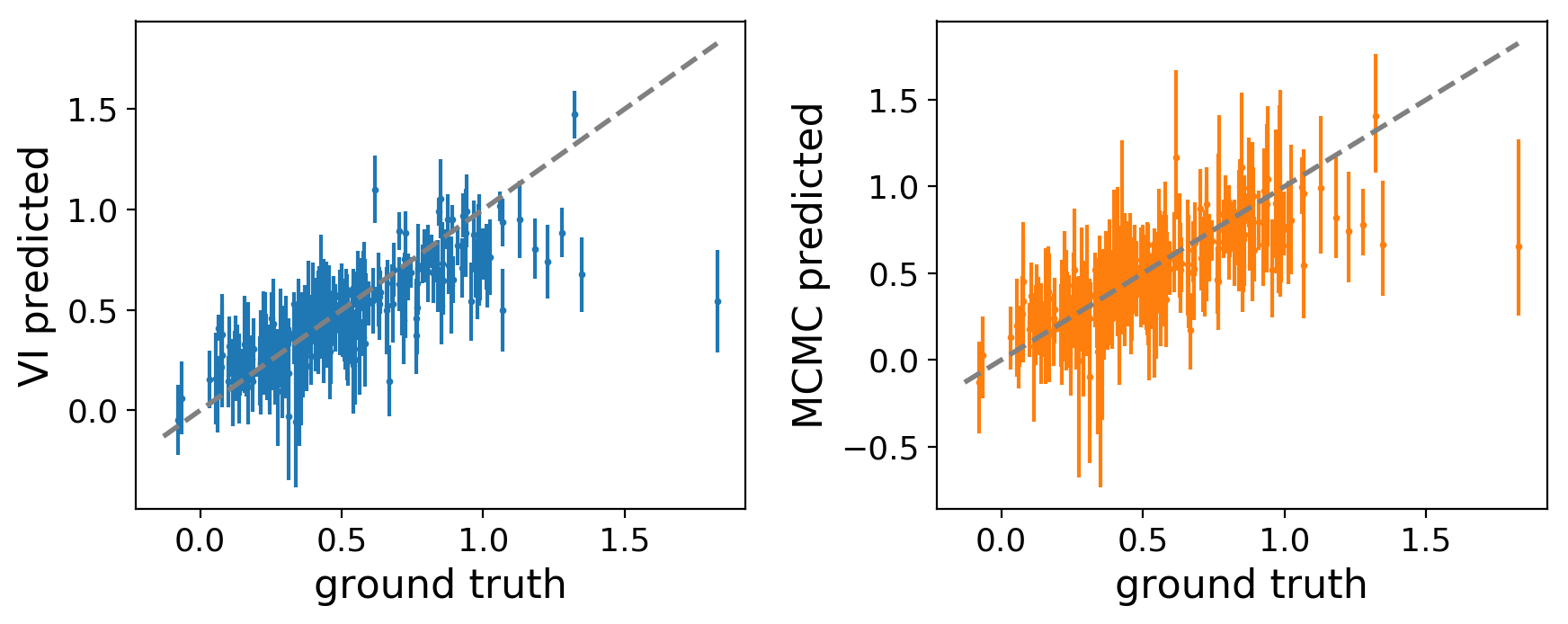}
  \caption{color r-i (galaxies)}
\end{subfigure}
\caption{VI and MCMC performance on real data from Stripe 82.  Each pair depicts VI (left, blue) and MCMC (right, orange), with the ground truth along the $x$-axis and the posterior distribution (showing equal-tailed 95.4\% credible intervals) along the $y$-axis.}
\label{qq-real}
\end{figure}

\begin{figure}[b]
\begin{subfigure}{.32\textwidth}
  \centering
  \includegraphics[width=\linewidth]{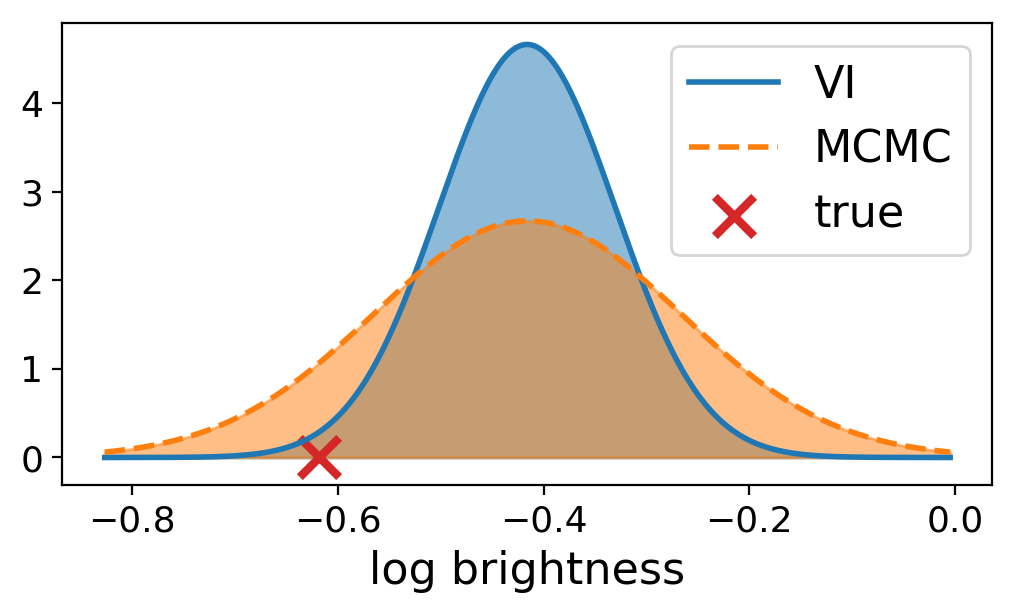}
  \caption{}
\end{subfigure}
\begin{subfigure}{.32\textwidth}
  \centering
  \includegraphics[width=\linewidth]{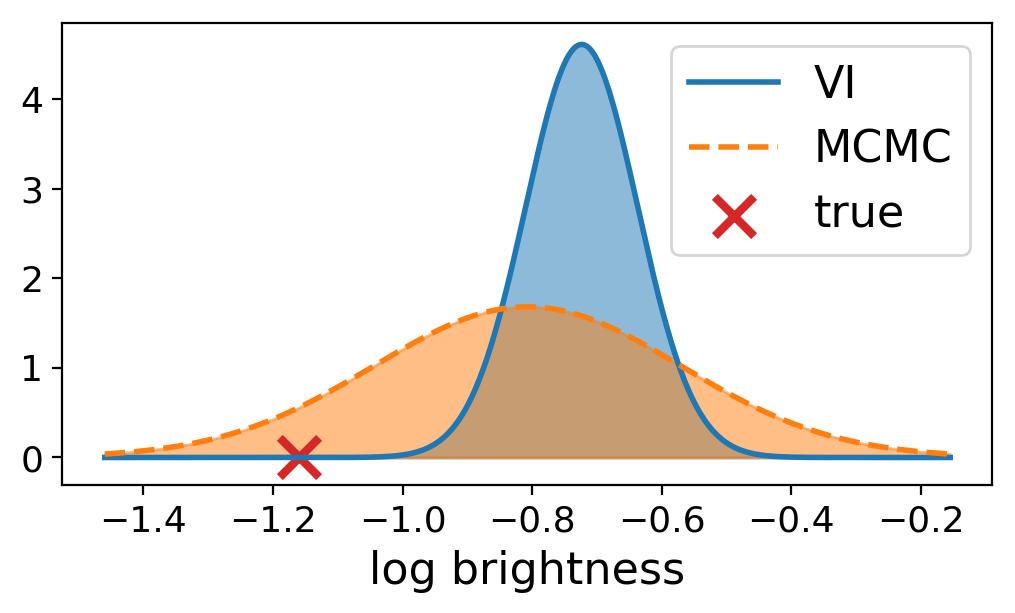}
  \caption{}
\end{subfigure}
\begin{subfigure}{.32\textwidth}
  \centering
  \includegraphics[width=\linewidth]{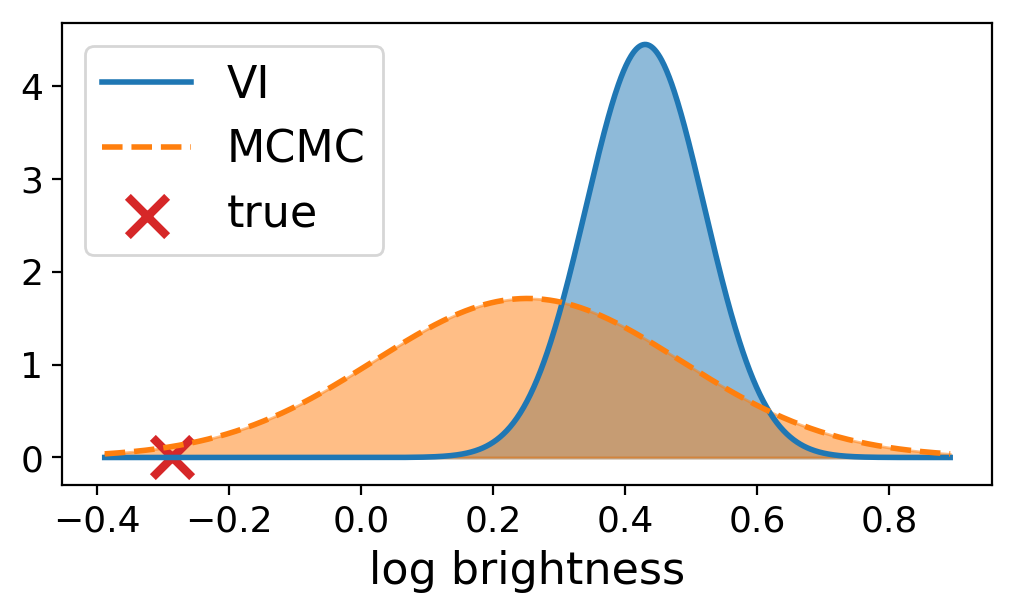}
  \caption{}
\end{subfigure}
  \caption{Comparison of posterior uncertainty for the flux of three light sources from Stripe 82 where the posterior mean is a poor prediction of the true parameter value. VI underestimates posterior uncertainty. MCMC assigns much greater posterior density to the true values.}
\label{uq-real}
\end{figure}

For SDSS images, MCMC had better calibrated uncertainty estimates, particularly for log flux (Figure~\ref{qq-real}, Figure~\ref{uq-real}, and Table~\ref{tab:calibration-s82}).
Recall that on the synthetic data, MCMC substantially outperformed VI at modeling uncertainty, producing empirical uncertainties that followed their theoretical distribution almost exactly (Table~\ref{tab:calibration-synth}).
On real data, uncertainty estimates for both MCMC and VI are worse than on synthetic data.
Model misspecification appears to have an effect on MCMC that is comparable to the effect of the bias introduced by the independence assumptions of the variational distribution.

Table~\ref{tab:err-s82} also shows that both MCMC and VI have lower error than Photo (previous work) on many metrics. It should be noted, however, that Photo does not make use of prior information, whereas both MCMC and VI do. For many downstream applications, something like Bayesian shrinkage (e.g., via corrections for Eddington bias, or use of default or empirical priors in a Bayesian setting) would first be applied to Photo's estimates---our comparison is not directly applicable for these applications. For the downstream application of selecting spectrograph targets, Photo's estimates are typically used without adjusting for prior information. For this application our results suggest that either our VI or our MCMC procedure may work better than Photo.
Hence, these results, though suggestive, do not conclusively establish that our method outperforms Photo.

\begin{table}
  \centering
  \scalebox{.85}{
  \begin{tabular}{l|lll|rlrlrl}
\toprule
{} &    MCMC &      VI &   Photo &              \multicolumn{2}{c}{Photo-VI} &             \multicolumn{2}{c}{Photo-MCMC} &                \multicolumn{2}{c}{VI-MCMC} \\
\midrule
direction   &   0.266 &   0.268 &   0.271 &  0.003 & \hspace{-1em}($\pm$ 0.011) &   0.004 & \hspace{-1em}($\pm$ 0.010) &   0.001 & \hspace{-1em}($\pm$ 0.002) \\
flux &   0.163 &   0.159 &   0.168 &  0.009 & \hspace{-1em}($\pm$ 0.013) &   0.005 & \hspace{-1em}($\pm$ 0.013) &  -0.005 & \hspace{-1em}($\pm$ 0.008) \\
color u-g  &   0.574 &   0.589 &   0.943 &  \textbf{0.417} & \hspace{-1em}($\pm$ 0.063) &   \textbf{0.428} & \hspace{-1em}($\pm$ 0.063) &   0.015 & \hspace{-1em}($\pm$ 0.008) \\
color g-r  &   0.146 &   0.146 &   0.293 &  \textbf{0.147} & \hspace{-1em}($\pm$ 0.020) &   \textbf{0.147} & \hspace{-1em}($\pm$ 0.019) &   0.0005 & \hspace{-1em}($\pm$ 0.003) \\
color r-i  &   0.096 &   0.097 &   0.175 &  \textbf{0.078} & \hspace{-1em}($\pm$ 0.010) &   \textbf{0.079} & \hspace{-1em}($\pm$ 0.010) &   0.001 & \hspace{-1em}($\pm$ 0.002) \\
color i-z  &   0.158 &   0.153 &   0.336 &  \textbf{0.184} & \hspace{-1em}($\pm$ 0.026) &   \textbf{0.179} & \hspace{-1em}($\pm$ 0.026) &  -0.005 & \hspace{-1em}($\pm$ 0.003) \\
galaxy profile    &   0.268 &   0.195 &   0.245 &  \textbf{0.050} & \hspace{-1em}($\pm$ 0.019) &  -0.023 & \hspace{-1em}($\pm$ 0.018) &  \textbf{-0.073} & \hspace{-1em}($\pm$ 0.015) \\
galaxy axis       &   0.115 &   0.146 &   0.219 &  \textbf{0.073} & \hspace{-1em}($\pm$ 0.012) &   \textbf{0.104} & \hspace{-1em}($\pm$ 0.012) &   \textbf{0.031} & \hspace{-1em}($\pm$ 0.005) \\
galaxy radius     &   0.572 &   0.692 &   1.274 &  0.582 & \hspace{-1em}($\pm$ 0.299) &   \textbf{0.701} & \hspace{-1em}($\pm$ 0.293) &   0.120 & \hspace{-1em}($\pm$ 0.067) \\
galaxy angle      &  19.32 &  19.54 &  20.39 &  0.838 & \hspace{-1em}($\pm$ 1.164) &   1.062 & \hspace{-1em}($\pm$ 1.165) &   0.225 & \hspace{-1em}($\pm$ 0.549) \\
\bottomrule
\end{tabular}

  }
  \vspace{0.5em}
  \caption{%
\textit{Left columns:} Mean absolute error on Stripe 82 data.
\textit{Right columns:}~Pairwise error differences for each pair of methods (and standard error).
Statistically significant differences appear in bold font.}
\label{tab:err-s82}
\end{table}

\begin{table}
\scalebox{.75}{
\begin{tabular}{lrrrr}
{} &  &  \textbf{VI} &  & \\
\toprule
{} & within 1/2 sd &  1 sd &  2 sd &  3 sd \\
\midrule
log flux &         0.12 & 0.21 & 0.39 & 0.58 \\
color u-g   &         0.25 & 0.44 & 0.75 & 0.89 \\
color g-r   &         0.25 & 0.48 & 0.76 & 0.91 \\
color r-i   &         0.22 & 0.41 & 0.72 & 0.87 \\
color i-z   &         0.27 & 0.51 & 0.81 & 0.94 \\
\bottomrule
\end{tabular}

}~
\scalebox{.75}{
\begin{tabular}{lrrrr}
{} &  &  \textbf{MCMC} &  & \\
\toprule
{} & within 1/2 sd &  1 sd &  2 sd &  3 sd \\
\midrule
log flux &         0.18 & 0.37 & 0.67 & 0.82 \\
color u-g   &         0.30 & 0.57 & 0.85 & 0.91 \\
color g-r   &         0.34 & 0.59 & 0.85 & 0.94 \\
color r-i   &         0.30 & 0.58 & 0.88 & 0.95 \\
color i-z   &         0.33 & 0.57 & 0.87 & 0.95 \\
\bottomrule
\end{tabular}

}
\vspace{.5em}
\caption{%
Proportion of light sources having posterior means found by VI (left) and MCMC (right) near the ground truth for SDSS images.
Credible interval widths match standard deviations as described in Table~\ref{tab:calibration-synth}.}
\label{tab:calibration-s82}
\end{table}

\subsection{Runtime comparison}
MCMC took approximately $1000\times$ longer in wall-clock time than VI to attain good results.
The implementations for MCMC and VI were both carefully optimized for speed, to make their runtimes comparable. In fact, the majority of runtime for MCMC was spent in code also used by VI, since the most computationally intensive calculations (across pixels) are shared by both the variational lower bound and the log likelihood function. This largely rules out ``implementation differences'' as an explanation for the disparity in runtime.

The same hardware was used for all timing experiments: a single core of an Intel Xeon E5-2698 v3 clocked at 2.30GHz.

Our MCMC experiments use a temperature schedule of length 200 for annealed importance sampling (AIS).
We repeated AIS 25 times to generate 25 independent estimates of the normalizing constant for each model.
We then ran each of these 25 independent posterior samples for 25 more slice sampling steps, generating 625 correlated samples.
For MCMC, the number of samples drawn scales linearly with runtime, presenting a speed/accuracy trade-off.
However, the quality of an MCMC posterior approximation is a function of the number of effectively independent samples~\citep{gelman2014bayesian}.
We measure the rate at which slice sampling is able to compute effectively independent samples for a single source ($52\times 52$ image patch).
For stars, we compute 0.225 effectively independent samples per second.
For galaxies, we compute 0.138 effectively independent samples per second.
VI is able to compute an approximate posterior distribution for one light source in 9 seconds, on average, for a region of sky imaged once in each of five filter bands. This runtime holds for either synthetic or SDSS data; runtime is largely determined by the number of pixels.

\subsection{Deblending}
\begin{figure}
	\begin{subfigure}{0.45\textwidth}
		\includegraphics[width=1.8in]{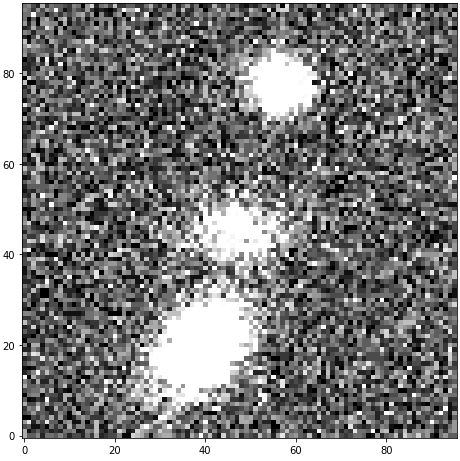}
		\subcaption{Two galaxies and one star. Their centers are on a line.}
		\label{three_sources_in_a_row}
	\end{subfigure}
	\hfill
	\begin{subfigure}{0.45\textwidth}
		\includegraphics[width=1.8in]{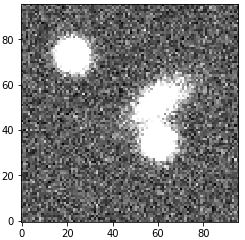}
		\subcaption{Two stars and one galaxy, all having 10-nanomaggy flux density.}
		\label{three_sources_two_overlap}
	\end{subfigure}
	\caption{Simulated astronomical images from GalSim.}
	\label{galsim_images}
\end{figure}

For the proposed model, overlapping light sources are not a special case requiring special processing logic.
Existing cataloging pipelines, on the other hand, invoke specialized ``deblending'' routines to deal with overlapping light sources, to avoid, for example, double counting photons.
In this section, we evaluate our procedure using simulated astronomical images from GalSim~\citep{rowe2015galsim}.
Using simulated rather than real data is particularly important for deblending experiments, because ground truth is particularly difficult to establish for overlapping light sources.
In contrast to our synthetic data (Section~\ref{sec:synth}), the simulated data is not drawn from our model, so there is the potential for model misfit.%

First, we consider images where three or more peaks in a blend appear in a straight line,
because this case was the ``single biggest failure mode'' for the deblending algorithm used by the Hyper Suprime-Cam (HSC) software pipeline~\citep{bosch2018hyper}.
To verify that this represents no special challenge to our model, we generated the astronomical image in Figure~\ref{three_sources_in_a_row}.
The correct r-band fluxes of the light sources, ordered from bottom to top, are 10 nanomaggies, 3 nanomaggies, and 3 nanomaggies.
Our VI procedure correctly classifies all three and determines that their respective flux densities are 9.98 nanomaggies, 2.90 nanomaggies, and 3.01 nanomaggies. The classifications are correct (assigning greater than 99\% probability to the truth), and mean galaxy angles are both within a few degrees of the truth.
We do not report the HSC pipline's estimation on this image because we could not get it to run without errors.

Second, we consider images with more severe blending and compare our algorithm to SExtractor~\citep{bertin1996sextractor}.
Unlike the SDSS and HSC pipelines, SExtractor is relatively straightforward to run on new data.
Recently released Python bindings make using it particularly straightforward~\citep{barbary2016sep}.
SExtractor is among the most used cataloging software today.

Figure~\ref{three_sources_two_overlap} shows a second simulated image we used for testing.
These light sources all have high flux density---10 nanomaggies each.
The approximate posterior mean recovered by our VI procedure assigns  9.87 nanomaggies, 9.95 nanomaggies, and 10.12 nanomaggies to these light sources.
SExtractor, on the other hand, estimates their flux densities to be 10.85 nanomaggies, 12.81 nanomaggies, and 14.91 nanomaggies.

\cite{melchior2018scarlet} propose a new deblending algorithm, called SCARLET, and report improvements over the HSC approach to deblending.
SCARLET appears at first glance to be quite different from our approach:
it is based on non-negative matrix factorization (NMF) rather than Bayesian statistics.
However, NMF algorithms can be cast as computing a maximum a posteriori (MAP) estimate under some assumptions on the distribution of the data and the factors~\citep{schmidt2009bayesian}, so SCARLET may have some similarity to what we propose.

\section{Bayesian inference at petascale}
\label{sec:at-scale}

Catalog inference is a ``big data'' problem that does not parallelize trivially.
This section introduces high-performance computing (HPC) to a statistics audience by describing large-scale runs of our variational inference procedure.
We construct a catalog from the entire 50-terabyte SDSS dataset. More importantly, we attain the computational efficiency needed to process the next generation of surveys, which will include $\mathcal O(100)$ petabytes of image data.

\subsection{Hardware}

Our test platform was the Cori supercomputer---currently ranked eighth in the global ``Top 500'' rankings~\citep{top500}.
Cori comprises 9,688 compute nodes connected by a high-speed network~\citep{cori}.
Each compute node has 112 GB of memory and one processor, an Intel Xeon Phi 7250, commonly referred to as ``Knights Landing.''
Though Knights Landing runs at only 1.4 GHz, it more than makes up for this relatively slow clock by executing many instructions in parallel during each clock cycle.
A single Knights Landing processor has 68 cores---physically distinct regions of the processor that execute instructions in parallel.
Each core simultaneously runs two hardware threads that appear to the operating system as separate cores.
A hardware thread executes batches of instructions twice per clock cycle: once on the ``up-tick'' and once on the ``down-tick.''
During each tick, a hardware thread may execute the same instruction on eight different 64-byte floating point numbers. This is known as single-instruction multiple-data (SIMD) parallelism.

\subsection{Efficient thread-level execution}
\label{thread}

Supercomputer programs are written almost exclusively in verbose languages like assembly, Fortran, C, and C\texttt{++}.
Many statisticians, however, prefer very high-level (VHL) languages like R and Python.
These languages often require $5\times$ to $10\times$ fewer lines of code to express the same algorithm.
Unfortunately, they also often run $10\times$, $100\times$, or even $1000\times$ slower than equivalent C code~\citep{juliabenchmarks}.
For high-performance computing, these languages are therefore limited to serving as ``glue'' code that connects libraries (e.g., BLAS, TensorFlow) that are implemented in more efficient languages.
In turn, writing code in two languages prevents many optimizations~\citep{bezanson2017julia}.

Our work uses the Julia programming language~\citep{bezanson2017julia} for the first time in an HPC setting.
Julia matches both the succinctness of scripting languages and the speed of C.
The ``hot spots'' in a Julia codebase, however, must be written carefully to attain C-like speed.

The process of tuning Julia code to run in an HPC setting is iterative.
It begins with profiling a typical execution of the code to find bottlenecks;
intuition about which lines of code are hotspots is a poor substitute for measurement.
Our first round of bottlenecks involved memory allocation, where the program requests that the operating system assign it more memory. We removed all these memory allocations from loops that contributed significantly to runtime by allocating the memory up front (i.e., ``pre-allocating'' memory).

The next round of bottlenecks was due to memory access: processors cannot execute instructions until data has been transferred from main memory to the processor's registers. A hardware thread may remain idle for approximately 200 clock cycles while fetching one number from main memory.
Memory-access bottlenecks need to be fixed on a case-by-case basis. The solution typically involves some reordering of the computation to enable better prefetching of data from main memory. In some cases, we save time by recomputing values rather than fetching them.

\subsection{Multi-node scaling}
In HPC, ``scalability'' refers to how a program's performance varies with the capacity of the hardware devoted to executing the program \citep{hager2010introduction}.
We assess scaling empirically in two ways.
First, we vary the number of compute nodes while keeping the amount of work constant per compute node (``weak scaling''); many compute nodes can solve a much larger problem.
Here the problem size is the area of the sky that we are constructing a catalog for.
Second, we vary the number of compute nodes while keeping the total job size constant (``strong scaling''); many compute nodes have to further subdivide the problem.
The two scaling metrics give different perspectives to inform predictions about how a particular supercomputer program will perform on future datasets, which may be much larger than any of the datasets used for testing.

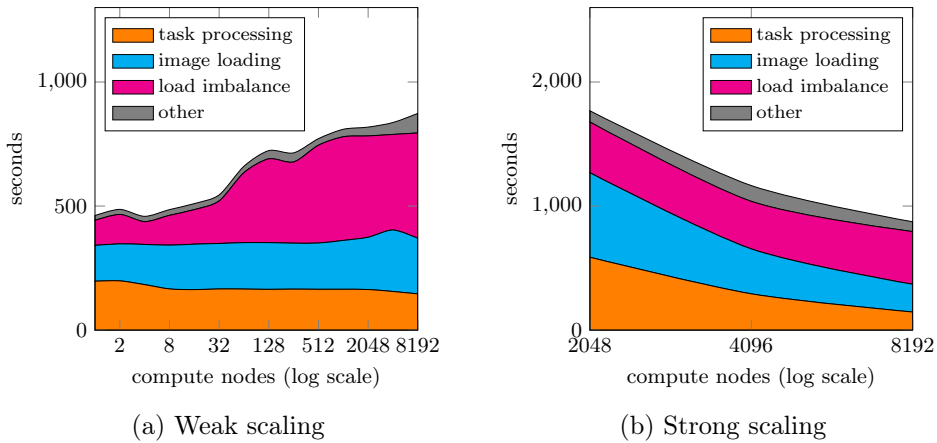
\begin{figure}[ht]
\begin{subfigure}{.48\textwidth}
\begin{tikzpicture}[scale=.9]
\begin{semilogxaxis}[
const plot,
smooth,
width=2.5in,
height=2.5in,
stack plots=y,
area style,
xlabel=compute nodes (log scale),
xtick={2,8,32,128,512,2048,8192},
xticklabels={2,8,32,128,512,2048,8192}, 
enlarge x limits=false,
log basis x={2},
log ticks with fixed point,
ymin=0,
ymax=1300,
ylabel=seconds,
legend pos=north west,
legend columns=1,
legend cell align={left},
legend style={font=\scriptsize}]
\addplot[fill=orange] table[x=nodes,y=opt_srcs] {figures/weak_scaling_GB.txt}
\closedcycle;
\addlegendentry{task processing}
\addplot[fill=cyan] table[x=nodes,y=load_wait] {figures/weak_scaling_GB.txt}
\closedcycle;
\addlegendentry{image loading}
\addplot[fill=magenta] table[x=nodes,y=wait_done] {figures/weak_scaling_GB.txt}
\closedcycle;
\addlegendentry{load imbalance}
\addplot[fill=gray] table[x=nodes,y=other] {figures/weak_scaling_GB.txt}
\closedcycle;
\addlegendentry{other}
\end{semilogxaxis}
\end{tikzpicture}
\caption{Weak scaling}
\label{fig:weak_scaling}
\end{subfigure}
\hfill
\begin{subfigure}{.48\textwidth}
\begin{tikzpicture}[scale=.9]
\begin{semilogxaxis}[
const plot,
smooth,
width=2.5in,
height=2.5in,
stack plots=y,
area style,
xtick={2048,4096,8192},
xticklabels={2048,4096,8192},
xlabel=compute nodes (log scale),
enlarge x limits=false,
log basis x={2},
log ticks with fixed point,
ymin=0,
ymax=2600,
ylabel=seconds,
legend pos=north east,
legend columns=1,
legend cell align={left},
legend style={font=\scriptsize}]
\addplot[fill=orange] table[x=nodes,y=opt_srcs]{figures/strong_scaling_GB.txt}
\closedcycle;
\addlegendentry{task processing}
\addplot[fill=cyan] table[x=nodes,y=load_wait]{figures/strong_scaling_GB.txt}
\closedcycle;
\addlegendentry{image loading}
\addplot[fill=magenta] table[x=nodes,y=wait_done]{figures/strong_scaling_GB.txt}
\closedcycle;
\addlegendentry{load imbalance}
\addplot[fill=gray] table[x=nodes,y=other]{figures/strong_scaling_GB.txt}
\closedcycle;
\addlegendentry{other}
\end{semilogxaxis}
\end{tikzpicture}
\caption{\textcolor{black}{Strong scaling}}
\label{fig:strong_scaling}
\vfill
\end{subfigure}
\caption{Scaling results. Load imbalance is due to the limited size of our study dataset---real datasets will be much larger. See text for additional discussion.}
\label{fig:scaling}
\end{figure}

Generally, it is harder to use more compute nodes efficiently. Ideal weak scaling is constant runtime as the number of compute nodes increases. Figure~\ref{fig:weak_scaling} shows instead that our runtime roughly doubles as the number of compute nodes increases from 1 to 8192. Ideal strong scaling is runtime that
drops by a factor of $1/c$ when the number of compute nodes grows by a factor of $c$.
Figure~\ref{fig:strong_scaling} shows instead that our runtime roughly halves as the number of compute nodes quadruples from 2048 to 8192.

Additionally, the scaling graphs break out runtime by component.
The \textit{image loading} component is the time taken to load images while worker threads
are idle. After the first task, images are prefetched in the background, so the majority of image loading time accrues up front.
Image loading time is constant in the weak scaling graph and proportional to the inverse of the number of nodes in the strong scaling graph---exactly what we want. We are not I/O bound even at high node counts.

The \textit{load imbalance} component is time when processes are idle because no tasks remain, but the job has not ended because at least one process has not finished its current task.
Both scaling graphs indicate that load imbalance is our primary scaling bottleneck.
Fortunately, the load imbalance is due to having only 4 tasks per process.
With at least $1000\times$ more data, the volume we expect from LSST, the load imbalance should become negligible.

The \textit{task processing} component is the main work loop.
It involves no network or disk I/O, only computation and shared memory access.
Because of this, task processing serves as a sanity check for both graphs: it should, and does, stay roughly constant in the weak scaling graph and vary in inverse proportion to the number of nodes in the strong scaling graph.

The \textit{other} component is everything else. It is always a small fraction of the total runtime.
It includes scheduling overhead, network I/O (excluding image loading), and writing output to disk.

\subsection{Peak performance}
To assess the peak performance that can be achieved for Bayesian inference at scale, we prepared a specialized configuration for performance measurement in which the processes synchronize after loading images, prior to task processing. We ran this configuration on 9568 Cori Intel Xeon Phi nodes, each running 17 processes of eight threads each, for a total of 1,303,832 threads. 57.8 TB of SDSS image data was loaded over a ten-minute interval. (Some regions were loaded multiple times, as prescribed by our algorithm.) The peak performance achieved was 1.54 PFLOP/s in double-precision. To the best of our knowledge, this experiment (conducted in May 2017) was the first time a supercomputer program in any language other than C, C\texttt{++}, Fortran, or assembly has exceeded one petaflop in double-precision.

\subsection{Complete SDSS catalog}
In a long-running job with 256 compute nodes, we constructed a preliminary astronomical catalog based on the entire SDSS. The catalog is 21 GB and contains 112 million light sources. Spot checking results gives us high confidence that distributed executions of our program give the same results as serial executions.

Our catalog contains the parameters of the optimal variational distribution---a vector with 44 single-precision floating point numbers for each light source.
We are considering both the FITS file format~\citep{wells1979fits} and the HDF5 file format~\citep{folk2011overview} for distributing future catalogs.
FITS is the standard format for astronomical images and catalogs, whereas the HDF5 format has better I/O speed and compression~\citep{price2015hdfits}.

\subsection{Future hardware}
In July, 2018, it was reported that Intel will discontinue development of the Xeon Phi line of processors~\citep{phieol}.
Future supercomputers will likely be based instead on the Xeon Scalable Family line of processors~\citep{intelroadmap} and the AMD Epyc~\citep{amdepyc}.
Both are ``many core'' processors having tens of cores, like the Xeon Phi, but they are clocked at a higher rate. Running efficiently on these processors should not require significant changes to our algorithm or to our Julia implementation. The Julia compiler and LLVM, on the other hand, may require optimizations to fully exploit the capabilities of these processors.

The next generation of supercomputers may also rely more on GPUs to attain exascale performance~\citep{summit}.
The variable size of imaged light sources makes SIMD parallelization across light sources somewhat challenging.
A different approach to parallelization may be advisable for astronomical cataloging on GPU-based clusters.

\section{Discussion}
\label{sec:discussion}

We introduced our work by identifying a limitation of existing cataloging pipelines: centroiding, deblending, photometry, star/galaxy separation, and incorporation of priors happen in distinct stages.
Uncertainty is typically not propagated between stages.
Any uncertainty estimates these pipelines produce are based on conditional distributions---that is, they are conditional on the output of the previous stages.

We developed a joint model of light sources' centers, colors, fluxes, shapes, and types (star/galaxy).
Whereas previous approaches to cataloging have been framed in algorithmic terms,
statistical formalisms let us characterize our inferences without ambiguity.
Statistical formalisms also make modeling assumptions transparent---whether the assumptions are appropriate ultimately depends on the downstream application.
We highlighted limitations of the model to guide further development.

A model is only useful when it can be applied to data.
We proposed two procedures: one based on MCMC and the other on VI.
Neither MCMC nor VI could be applied to our model without customization.
The need for problem-specific adjustments is a barrier to the broader adoption of both techniques.
With MCMC, for example, we went through several iterations before settling on slice sampling and AIS, including Metropolis-Hastings (MH) and reversible jump~\citep{green1995reversible}.
Compared to slice sampling, we found MH difficult to tune.
We found that reversible-jump MCMC required carefully constructed proposals to jump often enough between the star and galaxy models and was also difficult to tune.

VI required even more problem-specific customization.
Our VI techniques include the following: 1) approximating an integrand with its second-order Taylor expansion; 2) approximating the point-spread function with a mixture of Gaussians; 3) upper bounding the KL divergence between the color and a GMM prior; 4) limiting the variational distribution to a structured mean-field form; 5) limiting the variational distribution to point masses for some parameters; and 6) optimizing the variational lower bound with a variant of Newton's method rather than coordinate ascent. This final technique was particularly laborious, as it involved manually deriving and implementing both gradients and Hessians for a complicated function.

On synthetic data, MCMC was better at quantifying uncertainty, which is likely due to the restrictive form
of the variational distribution.
Additionally, MCMC provided uncertainty estimates for all latent random variables, whereas VI modeled some random variables as point masses---in effect recovering maximum a posteriori (MAP) estimates for them.
However, MCMC was approximately $1000\times$ slower than VI.

On real data, point estimates from VI were not always worse than point estimates from MCMC. Neither procedures' uncertainty estimates were perfectly calibrated for galaxies, suggesting some degree of model misspecification.
Imperfectly calibrated uncertainties can nonetheless be useful, e.g., for flagging particularly unreliable point estimates. Additionally, even if the uncertainties are ignored by downstream analyses, point estimates typically improve when uncertainty is modeled.
For questions requiring calibrated uncertainties, enhancing the galaxy model may help to reduce model misspecification.
Though the galaxy model we use---one with elliptical contours---is standard in astronomy, a more flexible galaxy model shows promise~\citep{regier2015deep}.

For spectrographic targeting, our current catalog should nonetheless
be an improvement over what came before: previously, uncertainty estimates and prior information were ignored. For analysis of subpopulations, however, we stress a key difference between our catalog and traditional astronomical catalogs: our catalog is based on prior information, whereas traditional catalogs are not.
Moreover, though our prior is accurate enough for large-scale cataloging and deblending, it likely is not accurate enough for a final scientific analysis of a particular subpopulation of light sources (e.g. the galaxies with an ``active galactic nucleus'').
For this use case, which is beyond the scope of our work, we suggest two approaches. First, a user can form a Laplace approximation, to ``remove'' our priors from the catalog and replace them with priors that are more suitable for their subpopulation. To facilitate, any catalog generated with our method should also contain parameters of the priors used to generate it.
Catalog users can then apply new priors directly to the catalog, without revisiting the image data; astronomers typically prefer to work with catalogs rather than images because catalogs are so much smaller.

We would prefer that users deal differently, however, with model misspecification that affects their analysis:
instead of trying to work around model misspecification, enhance our model.
Then, rerun our cataloging software, with the new model, on the images. This approach encourages users to adapt the statistical model and the priors to their needs and to treat the catalog as an intermediate data product~\citep{turon2010telescopes}.
While some work would be required to modify our model, the techniques we illustrate in this paper could still be followed to perform inference.
The MCMC procedure makes it particularly straightforward to make changes.

Because astronomical surveys are large (comprising terabytes of data now, and petabytes in the near future), scalability is of paramount concern.
We approximated the posterior for a large image dataset and demonstrated the scaling characteristics necessary to apply approximate Bayesian inference to hundreds of petabytes of images from the next generation of astronomical surveys.
Our optimization procedure found a stationary point, even though doing so required treating the full dataset as a single optimization problem.

Because of the relative ease of deriving and implementing MCMC, it could be a useful tool for trying different models and testing for misspecification prior to implementing VI. In some cases, it may be simpler to expend more computational resources to scale up the MCMC procedure than to implement VI. For the most computationally intensive problems, however, only VI can currently perform approximate inference.

\begin{supplement}[id=suppA]
  \sname{Supplement A}
  \stitle{Kullback-Leibler divergences}
  \sdatatype{kl.pdf}
  \sdescription{Formulas for KL divergences between common distributions that appear in the derivation of the variational lower bound.}
  \label{kl}
\end{supplement}

\bibliographystyle{imsart-nameyear}
\bibliographystyle{agsm}
\bibliography{references}

\section*{Supplement A: Kullback-Leibler divergences}
\label{vi_details}
This section gives the formulas for common Kullback-Leibler divergences that appear in the derivation of variational lower bound.

The KL divergence for $a_s$ is between two categorical distributions.
\begin{align}
D_{\mathrm{KL}}(q(a_s), p(a_s))
=\acute a_s\log\frac{\acute a_s}{\mathcal A}+\left(1-\acute a_s\right)\log\frac{1-\acute a_s}{1-\mathcal A}
\end{align}

The KL divergence for $u_s$ is between a point mass and a uniform distribution.
\begin{align}
D_{\mathrm{KL}}(q(u_s), p(u_s)) = \frac{1}{360 \times 180}
\end{align}

The KL divergence for $e_s^{angle}$ is between a point mass and a uniform distribution.
\begin{align}
D_{\mathrm{KL}}(q(e_s^{angle}), p(e_s^{angle})) = \frac{1}{180}
\end{align}

The KL divergence for $e_s^{radius}$ is between a point mass and a log-normal distribution.
\begin{align}
\begin{split}
D_{\mathrm{KL}}(q(e_s^{radius}), p(e_s^{radius}))=
&-\log 2\pi - \frac{1}{2}\log \mathcal E_2^{radius}\\
&- \frac{(\acute e_s^{radius} - \mathcal E_1^{radius})^2}{2\mathcal E_2^{radius}}
\end{split}
\end{align}

The KL divergence for $e_s^{profile}$ is between a point mass and a Beta distribution. Here $\mathrm B$ denotes the beta function.
\begin{align}
\begin{split}
D_{\mathrm{KL}}(q(e_s^{profile}), p(e_s^{profile}))
= &(\mathcal E_1^{profile} - 1)\log \acute e_s^{profile}\\
   &+ (\mathcal E_2^{profile} - 1) \log (1 - \acute e_s^{profile})\\
   &- \mathrm{B}(\mathcal E_1^{profile}, \mathcal E_2^{profile})
\end{split}
\end{align}

The KL divergence for $e_s^{axis}$ is between a point mass and a Beta distribution.
\begin{align}
\begin{split}
D_{\mathrm{KL}}(q(e_s^{axis}), p(e_s^{axis}))
= &(\mathcal E_1^{axis} - 1)\log \acute e_s^{axis}\\
&+ (\mathcal E_2^{axis} - 1) \log (1 - \acute e_s^{axis})\\
&- \mathrm{B}(\mathcal E_1^{axis}, \mathcal E_2^{axis})
\end{split}
\end{align}

The KL divergence for $r_s^{axis}$ is between two log-normal distributions.
\begin{align}
D_{\mathrm{KL}}(q(r_s | a_s = i), p(r_s | a_s = i)) =
\log\frac{\mathcal R_{i2}}{\hat r_{si}}
  +\frac{\hat r_{si} + \left(\acute r_{si}-\mathcal R_{i1}\right)^{2}}{2\mathcal R_{i2}}
  -\frac{1}{2}
\end{align}

\end{document}